\documentclass[useAMS,usenatbib]{mn2e}
\usepackage{multicol}
\usepackage{graphicx}
\usepackage{epstopdf}
\usepackage{epsfig}
\usepackage{longtable}
\usepackage{changepage}
\usepackage{setspace}
\usepackage[normalem]{ulem}
\def\mbh{$M_{\rm BH}$}
\def\mbhx{$M_{\rm BH,X}$}
\def\mbhd{$M_{\rm BH,dyn}$}

\def\chandra{{\it Chandra}}
\def\xmm{{\it XMM-Newton}}
\def\edd{$L_X/L_{\mathrm{Edd}}$}
\def\Gu{G$\mathrm{\ddot{u}}$ltekin}
\def\ergs{erg s$^{-1}$}
\def\gam{$\Gamma$}
\def\lx{$L_{\mathrm{X}}$}
\def\nbmc{$N_{\mathrm{BMC}}$}
\def\comp{Comptonization}
\def\GLedd{$\Gamma-L_{\rm X}/L_{\mathrm{Edd}}$}

\def\lo3{$L_{\rm [OIII]}$}

\begin{document}
\title[Constraining \mbh\ in low-accreting AGN]{Constraining black hole masses in 
low-accreting AGN using X-ray spectra}
\author[I. Jang et al.]{I. Jang,$^{1}$\thanks{E-mail: ijang2@masonlive.gmu.edu},
M. Gliozzi,$^{1}$ C. Hughes,$^{1}$ and L. Titarchuk$^{1}$\\
$^{1}$ School of Physics, Astronomy and Computational Sciences,
George Mason University, 4400 University Drive, Fairfax, VA 22030}

\date{Accepted . Received }

\pagerange{\pageref{firstpage}--\pageref{lastpage}} \pubyear{2014}

\maketitle

\label{firstpage}

\begin{abstract}
In a recent work we demonstrated that a novel X-ray scaling method, 
originally introduced for Galactic black holes (GBHs), can be reliably extended to 
estimate the mass of supermassive black holes accreting at a moderate to 
high level. Here we investigate the limits of applicability of this method to 
low-accreting active galactic nuclei, using a control sample with good-quality 
X-ray data and dynamically measured mass. 
For low-accreting AGNs  ($L_{\rm X}/L_{\rm Edd}\le10^{-4}$), because the basic 
assumption that the photon index positively correlates with the 
accretion rate no longer holds the X-ray scaling 
method cannot be used. Nevertheless, the inverse correlation 
in the \GLedd\ diagram, found in several low-accreting BHs and confirmed 
by this sample, can be used to constrain \mbh\ within a factor of $\sim10$ 
from the dynamically determined values. We provide a simple recipe to 
determine \mbh\ using solely X-ray spectral data, which can be used as 
a sanity check for \mbh\ determination based on indirect optical methods. 
\end{abstract}


\section{Introduction}
It is now widely accepted that black holes exist on very different scales,
with masses that range between 3--20 $M_\odot$ for stellar mass 
black holes (sMBHs) and $10^6-10^9~M_\odot$ for supermassive 
black holes (SMBHs) at the center of galaxies and in active
galactic nuclei (AGNs), with possibly intermediate black 
holes ($M_{\rm BH}=10^2-10^5~M_\odot$) 
whose nature is still a matter of debate. 

Recent studies 
have provided evidence for the presence of supermassive black holes at the center 
of virtually every galaxy with a prominent bulge and for the existence of
tight correlations between \mbh\ and several galaxy parameters such as the velocity 
dispersion or the mass of the bulge \citep{mago98,geb00a,fer00}. This bolsters
the importance and ubiquity of these systems in the universe and suggests that 
black hole and galaxy  growth may be closely related and that black holes 
are essential ingredients in the evolution of galaxies.

Black holes are fairly simple objects that are completely described 
by only three parameters, mass, spin, and charge, with the latter generally
negligible in astrophysical studies. However, since active BHs are not isolated 
systems but feed on the gas provided by a stellar companion or on the gas
accumulated at the center of galaxies, the dimensionless accretion rate $\dot m$  
(defined as $\dot{m}=L_{\rm bol}/L_{\rm Edd}$, where $L_{\rm bol}$ is the bolometric luminosity 
and $L_{\rm Edd}=1.3\times10^{38}\,M_{\rm BH}/M_{\odot}$ \ergs\ the Eddington luminosity) 
in Eddington units  should be counted as an additional basic parameter. 

The determination of \mbh\ is one of the most crucial tasks to shed light on 
accretion and ejection phenomena in both supermassive and stellar BHs, because 
\mbh\ sets the time and length scales in these systems, and may play a fundamental 
role in the formation and evolution of galaxies. 
The most direct way to determine \mbh\ is via dynamical methods. Under the 
assumption of Keplerian motion, a lower limit on the mass of the compact 
object can be determined in GBHs by measuring orbital period and velocity of the
visible stellar companion. Similarly, in nearby weakly active
or quiescent galaxies the estimate of \mbh\ can be inferred by directly modeling 
the dynamics of gas or stars in the 
vicinity of the black hole (e.g., Kormendy \& Richstone 1995; 
Magorrian et al. 1998). For highly active galaxies that show significant optical
variability, the estimate of \mbh\ relies upon the so-called ``reverberation mapping"
method, where the ``test particles" are represented by high-velocity 
gas clouds, whose dynamics are dominated by the BH gravitational force and are usually 
referred to as the broad-line region (BLR), since their radiation is dominated by broad 
emission lines \citep{peterson93}.

These direct methods are the most accurate and
reliable ways to constrain \mbh, but at the same time
have severe limitations: the methods applied to semi-quiescent galaxies
require the sphere of influence of the black hole to be resolved by the instruments,
and hence can be extended only to nearby objects. 
On the other hand, the reverberation
mapping technique requires significant resources and time, and cannot be applied to 
very luminous sources, whose variability is typically characterized by small-amplitude
flux changes occurring on very long timescales, or to sources without a detected BLR.
In order to circumvent these limitations, several secondary indirect 
methods have been developed \citep[see, e.g.,][]{vest09}.
 Most of them rely on some empirical relationship
between \mbh\ and different properties of the host galaxy or are based on
results obtained from the reverberation mapping such as the radius-luminosity
relationship \citep{kasp00}. However, the extension of these empirical 
relationships to systems with \mbh\ and $\dot m$ vastly different from the original
limited samples is still untested, and the majority of these techniques still 
requires a detected BLR, significantly restricting the number of possible AGNs
and the black hole mass range that can be studied.

In order to perform statistical studies of BHs and understand their evolutionary 
history and connection to their host galaxies, it is
important to explore alternative ways to determine the BH mass that are not 
dependent on the assumptions of optical-based methods. An important role in this field
may be played by X-ray-based methods, since the X-rays are nearly ubiquitous
in accreting BH systems regardless of their mass or accretion state, are less
affected by absorption than optical/UV emission, and are produced and reprocessed
in the vicinity of the BH thus closely tracking its activity.

Recently, \citet{shapo09} developed a new X-ray scaling method 
to determine \mbh\ for GBHs. This method is based on the positive
correlation between X-ray photon index $\Gamma$ (which is generally considered as a
reliable indicator of the accretion state of the source; 
see e.g., Shemmer et al. (2007, 2008); Esh et al. (1997)) and the source brightness,
parameterized by the normalization of the Bulk Motion Comptonization (BMC) model. 
The self-similarity of this spectral trend, which is observed in different GBHs 
during different outbursts events, makes it possible to estimate \mbh\ in any GBHs 
by scaling the dynamically-constrained value of \mbh\ of a GBH considered as a 
reference source.
 
With the assumption that AGNs follow the same spectral evolution
as GBHs (although on much longer timescales that cannot be directly probed), 
in our recent work, we tested whether this novel X-ray scaling method could be
extended to supermassive BHs. To this end, we utilized a sample of AGNs
with good X-ray data and whose BH mass had been already determined via reverberation
mapping. The results (on average the \mbh\ values determined with this method are 
within a factor of 2-3 from the reverberation mapping values) demonstrate that this 
method is reliable and robust for 
BH systems accreting at moderate and high rate ($\dot m \geq$ 1\%) and
can be used to determine BH masses at any scale \citep{glio11}. 

The presence of a positive correlation between \gam\ and \lx\ 
(which is at the basis of the X-ray scaling method) has been observed in 
highly and moderately accreting BH systems at all scales for several decades. 
For example, a spectral steepening as the source brightens is 
consistently observed in the evolution between canonical spectral states in GBHs 
(e.g., Esin et al. 1997; Homan et al. 2001, and references therein). A similar 
behavior has been also 
observed in individual and samples of AGNs (e.g., Markowitz \& Edelson 2001; 
Papadakis et al. 2002; Shemmer et al. 2008). 
On the other hand, in the very low accreting regime ($\dot{m}\ll1\%$) 
convincing evidence of a \gam\ $-$ \lx\ (or \edd) 
anti-correlation has been revealed only recently 
(e.g., Wu \& Gu 2008; Constantin et al. 2009; Gu \& Cao 2009; 
Younes et al. 2011; \Gu\ et al. 2012; see however Trump et al. 2011 
for an alternative view).

Here, we want to investigate the limits of applicability of this method to low-accreting
BH systems, by applying it to a sample of low-accreting AGNs, which possess
good-quality X-ray data (either from \chandra\ or \xmm\ satellites) and with \mbh\ determined
from direct dynamical methods. In principle, since the direct 
\gam\ $-$ \lx\ correlation is the foundation of the scaling 
method and since at very low accreting levels no positive
correlation is observed, it is expected that at a certain
threshold value of $\dot{m}$ the X-ray scaling method should break down. Nevertheless, it is 
important to test if this break down actually occurs (this would provide indirect support 
to the foundation of the method, i.e., the self-similar spectral behavior of BHs at 
all scales) and at which value of $\dot m $ does it occur. 

The description of the sample and the data reduction are provided in Section 2
and 3, respectively. The X-ray spectral analysis is performed in Section 4. 
In Section 5, we apply the X-ray scaling method 
and show that the overall
spectral behavior of our sample is consistent with an anti-correlation in the 
$\Gamma-L_X/L_{\rm Edd}$ plot, which provides an alternative way to 
constrain \mbh. The main results and their implications are summarized and discussed
in Section 6.

\section{Description of the sample}
In order to test X-ray-based methods to determine 
\mbh\ in low-accreting AGNs, we need to select objects 
that fulfill the following criteria: 1) they must have a direct and robust 
estimate of \mbh; 2) they must possess good-quality X-ray data; and 3)
they must accrete at low level, \edd\ $\ll1$\%. 

Our sample contains a total of 53 low-luminosity AGNs (LLAGNs) with 
bolometric luminosity less 
than $10^{42}$ \ergs\ (Ho et al. 2001) and whose black hole mass 
has been determined via dynamical methods.  
The physical properties of the sources are listed in Table \ref{tab1}, 
in which column (1) provides the source name, columns (2) and (3) 
right ascension and declination, (4) the \mbh\ value via the dynamical 
method, (5) the redshift-independent distance 
from the NSAS/IPAC Extragalactic Database (NED), (6) the $H_{\alpha}/H_{\beta}$ ratio, 
(7) the flux of the narrow component of [OIII]$\lambda 5007$, 
(8) the radio luminosity, 
(9) the AGN class (S=Seyfert galaxies, L=LINERs), 
and (10) the galaxy class. The narrow-line emission and types were gathered from 
the Palomar survey \citep{ho97} unless stated otherwise. Based 
on the optical classification, this sample comprises 16 LINERs, 17 Seyfert 
galaxies ranging from type 1 to type 2, and 20 AGNs that are not optically 
classified. 

\begin{table*}
\begin{adjustwidth}{-0.3 in}{}
\footnotesize
\caption{The sample of LLAGNs. Optical and radio properties}
\label{tab1}
\begin{center}
\footnotesize
\begin{tabular}{llclcccccc} 
\hline        
\hline
\noalign{\smallskip}
\multicolumn{1}{c}{Source Name} & \multicolumn{1}{c}{RA} & DEC & \multicolumn{1}{c}{$\log(M_{BH}/M_{\odot})$} &
\multicolumn{1}{c}{D (Mpc)} & \multicolumn{1}{c}{H$_{\alpha}$/H$_{\beta}$} &
\multicolumn{1}{c}{$\log(F_{[\rm OIII]})$} & $\log(L_{\rm 6 cm})$ &
\multicolumn{1}{c}{AGN Class} & \multicolumn{1}{c}{Optical Class}\\
\multicolumn{1}{c}{(1)} & \multicolumn{1}{c}{(2)} & (3) & \multicolumn{1}{c}{(4)} &
\multicolumn{1}{c}{(5)} & \multicolumn{1}{c}{(6)} &
\multicolumn{1}{c}{(7)} & (8) &
\multicolumn{1}{c}{(9)} & \multicolumn{1}{c}{(10)}\\
\hline
\noalign{\smallskip}
IC 1459 & 22 : 57 : 10.6 & -36 : 27 : 44 & $ 9.44 \pm 0.20 ^1$ & 29.2 & & & 39.76 & L/R  & E3 $^f$ \\
IC 4296 & 13 : 36 : 39.0 & -33 : 57 : 57 & $ 9.13 \pm 0.07 ^2$ & 62.2 & & & 38.59 & L  & E1 $^g$ \\
NGC 221 & 00 : 42 : 41.8 & 40 : 51 : 55 & $ 6.49 \pm 0.09 ^3$ & 0.81 & & & 33.3 & & E2  \\
NGC 224 & 00 : 42 : 44.3 & 41 : 16 : 09 & $ 8.17 \pm 0.16 ^4$ & 0.84 & & & 32.14 & L  & SA(s)b  \\
NGC 821 & 02 : 08 : 21.1 & 10 : 59 : 42 & $ 7.63 \pm 0.16 ^5$ & 22.4 & & &  & & E6  \\
NGC 1023 & 02 : 40 : 24.0 & 39 : 03 : 48 & $ 7.66 \pm 0.04 ^6$ & 9.82 & & &  & & SB(rs)0-  \\
NGC 1068 & 02 : 42 : 40.7 & 00 : 00 : 48 & $ 6.93 \pm 0.02 ^7$ & 10.1 & 5.29  & -10.46  & 39.18 & S2  & (r)SA(rs)b  \\
NGC 1300 & 03 : 19 : 41.1 & -19 : 24 : 41 & $ 7.85 \pm 0.29 ^8$ & 22.6 & & &  & & SB(rs)bc  \\
NGC 1399 & 03 : 38 : 29.1 & -35 : 27 : 03 & $ 8.71 \pm 0.06 ^9$ & 19.4 & & &  & S2 $^b$ & E1  \\
NGC 2748 & 09 : 13 : 43.0 & 76 : 28 : 31 & $ 7.67 \pm 0.50 ^8$ & 21.0 & 6.11  & -15.00  &  & H  & Sabc  \\
NGC 2778 & 09 : 12 : 24.4 & 35 : 01 : 39 & $ 7.21 \pm 0.32 ^5$ & 38.1 & & &  & & E2  \\
NGC 2787 & 09 : 19 : 18.6 & 69 : 12 : 12 & $ 7.64 \pm 0.05 ^{10}$ & 7.48 & 1.89  & -13.93  & 37.22 & L  & SB(r)0+  \\
NGC 3031 & 09 : 55 : 33.2 & 69 : 03 : 55 & $ 7.90 \pm 0.09 ^{11}$ & 3.65 & 3.15  & -12.65  & 36.82 & L $^b$ & SA(s)ab  \\
NGC 3115 & 10 : 05 : 14.0 & -07 : 43 : 07 & $ 8.98 \pm 0.18 ^{12}$ & 9.68 & & &  & S  & SA0- sping  \\
NGC 3227 & 10 : 23 : 30.6 & 19 : 51 : 54 & $ 7.18 \pm 0.23 ^{13}$ & 21.1 & 2.9  & -12.02  & 37.72 & S1.5  & SAB(s)a pec  \\
NGC 3245 & 10 : 27 : 18.4 & 28 : 30 : 27 & $ 8.35 \pm 0.11 ^{14}$ & 27.4 & 4.76  & -13.42  & 36.98 & L  & SA(r)0?  \\
NGC 3377 & 10 : 47 : 42.3 & 13 : 59 : 09 & $ 8.06 \pm 0.16 ^5$ & 11.3 & & -10.20 $^c$ &  & & E5+  \\
NGC 3379 & 10 : 47 : 49.6 & 12 : 34 : 54 & $ 8.09 \pm 0.25 ^{15}$ & 12.6 & & -14.16  &  & L  & E1  \\
NGC 3384 & 10 : 48 : 16.9 & 12 : 37 : 45 & $ 7.25 \pm 0.04 ^5$ & 10.8 & & -10.60 $^c$ &  & & SB(s)0-  \\
NGC 3585 & 11 : 13 : 17.1 & -26 : 45 : 17 & $ 8.53 \pm 0.12 ^{16}$ & 20.2 & & &  & & S0  \\
NGC 3607 & 11 : 16 : 54.6 & 18 : 03 : 06 & $ 8.08 \pm 0.15 ^{16}$ & 22.8 & 5.56  & -13.26  &  & S2 $^b$ & SA(s)0:  \\
NGC 3608 & 11 : 16 : 58.9 & 18 : 08 : 55 & $ 8.32 \pm 0.17 ^5$ & 23.1 & & -14.16  &  & L  & E2  \\
NGC 3998 & 11 : 57 : 56.1 & 55 : 27 : 13 & $ 8.37 \pm 0.43 ^{17}$ & 19.4 & 4.72  & -13.13  & 38.03 & L $^b$ & SA(r)0?  \\
NGC 4026 & 11 : 59 : 25.2 & 50 : 57 : 42 & $ 8.33 \pm 0.11 ^{16}$ & 11.7 & 3.4  & -10.95  &  & & (R')SAB(rs)ab:  \\
NGC 4151 & 12 : 19 : 23.2 & 05 : 49 : 31 & $ 7.65 \pm 0.05 ^{18}$ & 3.89 & & & 38.2 & S1.5  & SA0 spin  \\
NGC 4258 & 12 : 18 : 57.5 & 47 : 18 : 14 & $ 7.58 \pm 0.00 ^{19}$ & 7.59 & 3.94  & -12.98  & 36.03 & S1.9  & SAB(s)bc  \\
NGC 4261 & 12 : 19 : 23.2 & 05 : 49 : 31 & $ 8.74 \pm 0.09 ^{20}$ & 24.0 & 4.9  & -13.43  & 39.21 & L  & E2+  \\
NGC 4278 & 12 : 20 : 06.8 & 29 : 16 : 51 & $ 9.20 \pm 0.00 ^{21}$ & 10.0 & 2.5  & -13.17  & 37.91 & L  & E1+  \\
NGC 4291 & 12 : 20 : 18.2 & 75 : 22 : 15 & $ 8.51 \pm 0.34 ^5$ & 31.2 & & &  & & E  \\
NGC 4303 & 12 : 21 : 54.9 & 04 : 28 : 25 & $ 6.65 \pm 0.35 ^{22}$ & 12.2 & 3.92  & -12.97  & 38.46 & S2 $^b$ & SAB(rs)bc  \\
NGC 4342 & 12 : 23 : 39.0 & 07 : 03 : 14 & $ 8.56 \pm 0.19 ^{23}$ & 16.8 & & &  & & S0  \\
NGC 4374 & 12 : 25 : 03.7 & 12 : 25 : 04 & $ 9.18 \pm 0.23 ^{24}$ & 17.5 & 4.68  & -13.46  & 38.77 & S2  & E1  \\
NGC 4395 & 12 : 25 : 48.8 & 33 : 32 : 49 & $ 5.04 \pm 0.00 ^{25}$ & 4.83 & 2.13  & -12.46  & 35.56 & S1.8  & SA(s)m:  \\
NGC 4459 & 12 : 29 : 00.0 & 13 : 58 : 42 & $ 7.87 \pm 0.08 ^{10}$ & 16.6 & 3.24  & -14.64  &  & L  & SA(r)0+  \\
NGC 4473 & 12 : 29 : 48.9 & 13 : 25 : 46 & $ 8.11 \pm 0.35 ^5$ & 15.2 & & &  & & E5  \\
NGC 4486 & 12 : 30 : 49.4 & 12 : 23 : 28 & $ 9.56 \pm 0.13 ^{26}$ & 15.9 & 4.29  & -12.97  & 39.83 & L  & E0+pec  \\
NGC 4486A & 12 : 30 : 57.7 & 12 : 16 : 13 & $ 7.13 \pm 0.15 ^{27}$ & 15.1 & & &  &    E2  \\
NGC 4564 & 12 : 36 : 27.0 & 11 : 26 : 21 & $ 7.84 \pm 0.05 ^5$ & 15.8 & & &  & & E  \\
NGC 4594 & 12 : 39 : 59.4 & -11 : 37 : 23 & $ 8.76 \pm 0.41 ^{28}$ & 13.7 & 3.37  & -12.46 $^a$ & 37.89 & Sy1.9 $^b$ & SA(s)a spin  \\
NGC 4596 & 12 : 39 : 55.9 & 10 : 10 : 34 & $ 7.92 \pm 0.16 ^{11}$ & 16.8 & 2.12  & -15.35  &  & L  & SB(r)0+  \\
NGC 4649 & 12 : 43 : 40.0 & 11 : 33 : 10 & $ 9.33 \pm 0.12 ^5$ & 14.0 & & & 37.45 & & E2  \\
NGC 4697 & 12 : 48 : 35.9 & -05 : 48 : 03 & $ 8.29 \pm 0.04 ^5$ & 20.9 & & -15.90 $^d$ &  & & E6  \\
NGC 4742 & 12 : 51 : 48.0 & -10 : 27 : 17 & $ 7.18 \pm 0.15 ^{29}$ & 15.5 & & &  & & E $^h$ \\
NGC 4945 & 13 : 05 : 27.5 & -49 : 28 : 06 & $ 6.15 \pm 0.18 ^{30}$  & 4.50 & & & 38.17 & S $^b$ & \\
NGC 5077 & 13 : 19 : 31.7 & -12 : 39 : 25 & $ 8.90 \pm 0.22 ^{31}$ & 39.8 & 2.89  & -13.82  &  & L  & E3+  \\
\hline
\end{tabular}
\end{center}
\footnotesize
\begin{itemize}
\item[]
\textbf{\mbh\ Reference:} 
(1) Cappellari et al. 2002; (2) Dalla Bont$\grave{\rm a}$ et al. 2009; (3) Verolme et al. 2002; 
(4) Bender et al. 2005; (5) Gebhardt et al. 2000a; (6) Bower et al. 2001; 
(7) Lodato \& Bertin 2003; (8) Atkinson et al. 2005; (9) Gebhardt et al. 2007; 
(10) Sarzi et al. 2001; (11) Devereux et al. 2003; (12) Emsellem et al. 1999; 
(13) Hicks et al. 2008; (14) Barth et al. 2001; (15) Gebhardt et al. 2000b; 
(16) G$\ddot{\rm u}$ltekin et al. 2009a; (17) de Francesco et al. 2006; 
(18) Onken et al. 2007; (19) Herrnstein et al. 2005; (20) Ferrarese et al. 1996; 
(21) Cardullo et al. 2009; (22) Pastorini et al. 2007; (23) Cretton \& van den Bosch 1999; 
(24) Bower et al. 1998; (25) Haim et al. 2012; (26) Macchetto et al. 1997; 
(27) Nowak et al. 2007; (28) Kormendy 1988; 
(29) listed as M. E. Kaiser et al. 2002 in preparation in Tremaine et al. 2002 but never published; 
(30) Greenhill et al. 1997; (31) de Francesco et al. 2008; (32) Silge et al. 2005; 
(33) Capetti et al. 2005; (34) Ferrarese \& Ford 1999; (35) van der Marel \& van den Bosch 1998; 
(36) Wold et al. 2006\\
\end{itemize}
\end{adjustwidth}
\end{table*}

\begin{table*}
\footnotesize
\begin{adjustwidth}{-0.3 in}{}
Table 1 Continued $-$ The sample of LLAGNs. Optical and radio properties
\begin{center}
\footnotesize
\begin{tabular}{llclcccccc} 
\hline        
\hline
\noalign{\smallskip}
\multicolumn{1}{c}{Source Name} & \multicolumn{1}{c}{RA} & DEC & \multicolumn{1}{c}{$\log(M_{BH}/M_{\odot})$} &
\multicolumn{1}{c}{\it z} & \multicolumn{1}{c}{H$_{\alpha}$/H$_{\beta}$} &
\multicolumn{1}{c}{$\log(F_{[\rm OIII]})$} & $\log(L_{\rm 6cm})$ &
\multicolumn{1}{c}{AGN Class} & \multicolumn{1}{c}{Optical Class}\\
\hline
\noalign{\smallskip}
NGC 5128 & 13 : 25 : 27.6 & -43 : 01 : 09 & $ 8.48 \pm 0.04 ^{32}$ & 4.09 & 3.72 $^a$ & -13.15 $^e$ & 39.85 & S2  & \\
NGC 5252 & 13 : 38 : 15.9 & 04 : 32 : 33 & $ 9.00 \pm 0.34 ^{33}$ & 99.3 & 3.72 $^a$ & -12.41  & 39.05 & S2 $^b$ & S0  \\
NGC 5576 & 14 : 21 : 03.7 & 03 : 16 : 16 & $ 8.26 \pm 0.09 ^{16}$ & 25.8 & & &  & & E3  \\
NGC 5845 & 15 : 06 : 0.80 & 01 : 38 : 02 & $ 8.46 \pm 0.22 ^5$ & 24.1 & & &  & & E3  \\
NGC 6251 & 16 : 32 : 32.0 & 82 : 32 : 16 & $ 8.78 \pm 0.15 ^{37}$ & 97.6 & 15.1 $^a$ & -11.90  & 41.01 & S2  & E1  \\
NGC 7052 & 21 : 18 : 33.0 & 26 : 26 : 49 & $ 8.60 \pm 0.22 ^{38}$ & 56.7 & & & 39.43 & & E3  \\
NGC 7457 & 23 : 00 : 59.9 & 30 : 08 : 42 & $ 6.61 \pm 0.17 ^5$ & 12.3 & & -16.18  &  & S  & SA(rs)0-?  \\
NGC 7582 & 23 : 18 : 23.5 & -42 : 22 : 14 & $ 7.74 \pm 0.10 ^{39}$ & 22.2 & 7.6 $^a$ & -11.35  & 38.55 & {\bf S2} $^b$ & Sbab  \\
\hline
\end{tabular}
\end{center}
\footnotesize
\begin{itemize}
\item[]
\textbf{Note $-$} 
AGN Class $-$ S: Seyfert Galaxies, L: LINERs\\
$^a$ Bassani et al. 1999;  $^b$V\'eron-Cetty \& V\'eron 2006;  
$^c$Ciardullo et al. 1989; $^d$ M\'endez et al. 2005;  
$^e$ Walsh et al. 2012; $^f$ Fabbiano et al. 2003;  
$^g$Younis et al. 1985; $^h$ Naim et al. 1995
\end{itemize}
\end{adjustwidth}
\end{table*}

The dynamical mass measurements are based on high spatial resolution 
stellar velocity measurements \citep{gh08,gu09}, gas dynamic measurements \citep{ba01}, 
and maser measurements \citep{mi95}.
All sources have been observed with \chandra\ and most have also 
\xmm\ data. 
\chandra\ with its unsurpassed spatial resolution is ideal to disentangle 
the different X-ray components, whereas \xmm\ with its large collecting 
area and consequent higher sensitivity provides tighter constraints on 
the X-ray spectral parameters for isolated point-like sources. 

\section[]{Data reduction}
\subsection{\chandra\ observations}
In the \chandra\ archive, for each source we selected 
the observations with the longest exposures and the smallest 
pointing offset with respect to the nominal position of the AGN.
Most of the observations are the same as those analyzed by \cite{gu09}.
Adding six new sources observed more recently by \cite{gu12} and 
2 described by \cite{GC09}, the total number of sources with 
\chandra\ data is 53. 

The data reduction was carried out homogeneously for each source
as described below. 
Source's spectra and light curves were extracted from circular regions with 
a radius of $1"-2"$, and their background from nearby source-free 
circular regions with a radius of $10"-20"$. 
Sometimes, extraction regions were extended to $\sim3"$ for brighter sources.
The data reduction followed the standard pipeline, using \chandra\
data reduction software package version CIAO 4.4, and the nuclear source 
regions were confirmed after  
running {\it wavdetect} \citep{fr02}.
Background flares were cut above $3\sigma$ from the mean value.
The positions of source and background were given 
as input to the {\it specextract} tool to create 
the response matrix file (RMF) and ancillary response file (ARF).

\subsection{\xmm\ Observations}
36 sources (nearly 70\% of the \chandra\ sample) also possess \xmm\ data. 
We performed the data reduction following the standard procedures 
of Science Analysis System (SAS) version 12.0.1. 
We only selected good X-ray event (``FLAG=0") with 
patterns $0-4$ and $0-12$ for pn and MOS, respectively.
The positions inferred from \chandra\ were used as center 
of the extraction regions with a radius of $\sim10"$ 
or larger for brighter and extended sources.
When the nuclear source was located at the edge of a CCD or between two CCDs, 
we utilized the second longest \xmm\ observation. 
The background regions  were chosen in nearby empty spaces on 
the same CCD for both pn and MOS cameras. 
We used \chandra\ images and spectra to account for the presence of additional 
X-ray components (e.g., diffuse emission, jet-like structures, 
off-nuclear sources) in the \xmm\ extraction regions.
The SAS {\it rmfgen} and {\it argen} task were used to generate RMF and ARF files, respectively.
Of the 36 sources observed at least once by \xmm, 
33 of them have sufficient statistics for a meaningful spectral analysis. 

\subsection{Image Inspection}
Since the Point Spread Function (PSF) of \xmm\ EPIC cameras 
does not allow one to firmly distinguish between point-like and extended emission in 
low-luminosity 
sources, a systematic comparison of \xmm\ and \chandra\ images was done to 
investigate the presence of additional X-ray components 
in the extraction region.

To this end, we overlapped \chandra\ image contours 
on the corresponding \xmm\ images. 
From the \chandra\ images, 
we also measured source counts  
using different extraction regions (with radii of $10"$ and $20"$), in order to estimate 
the contribution of off-nuclear components encompassed by the 
larger extraction region of \xmm.   

\begin{figure*}
\begin{center}
\includegraphics[scale=0.25]{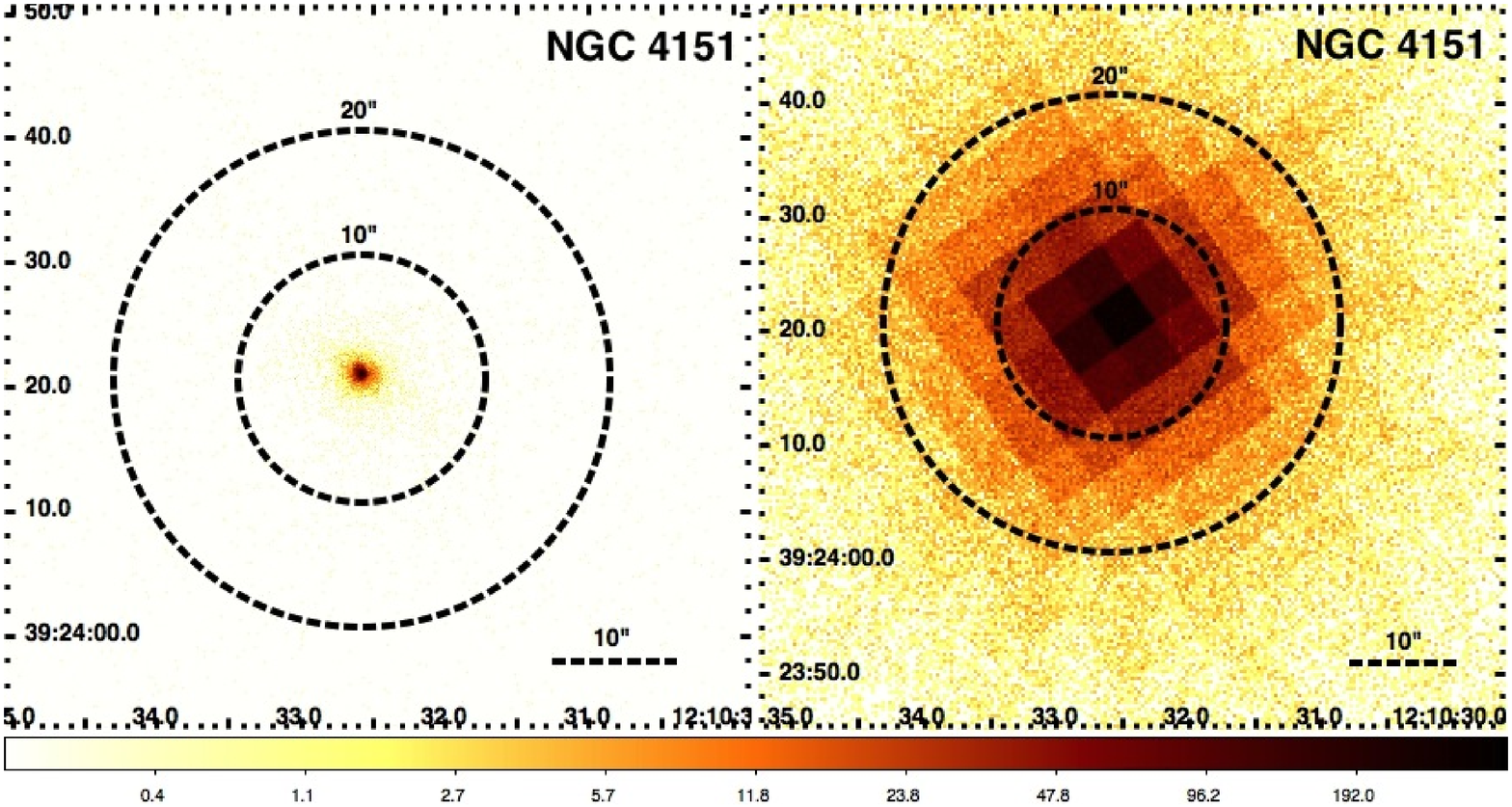}
\includegraphics[scale=0.18]{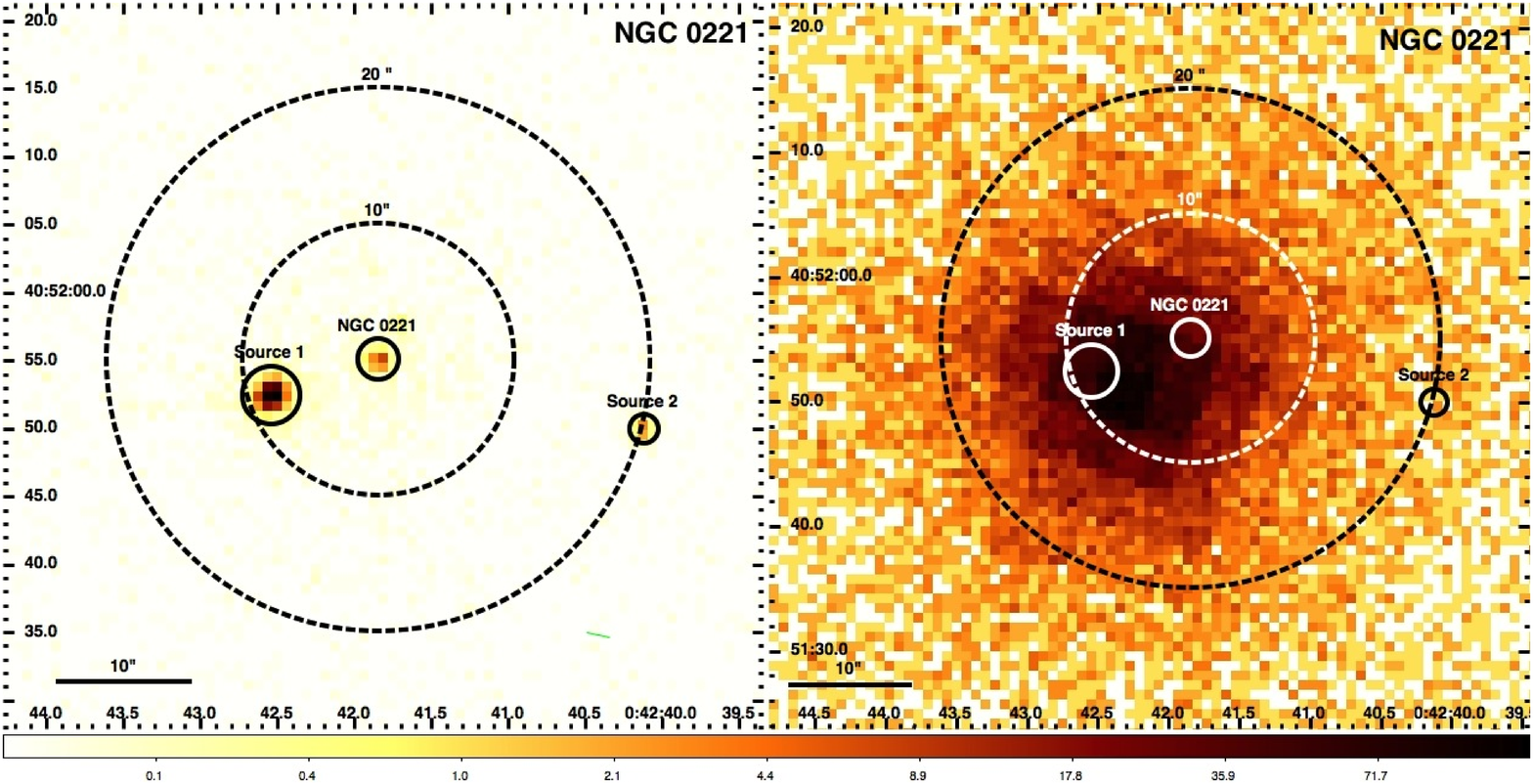}
\includegraphics[scale=0.25]{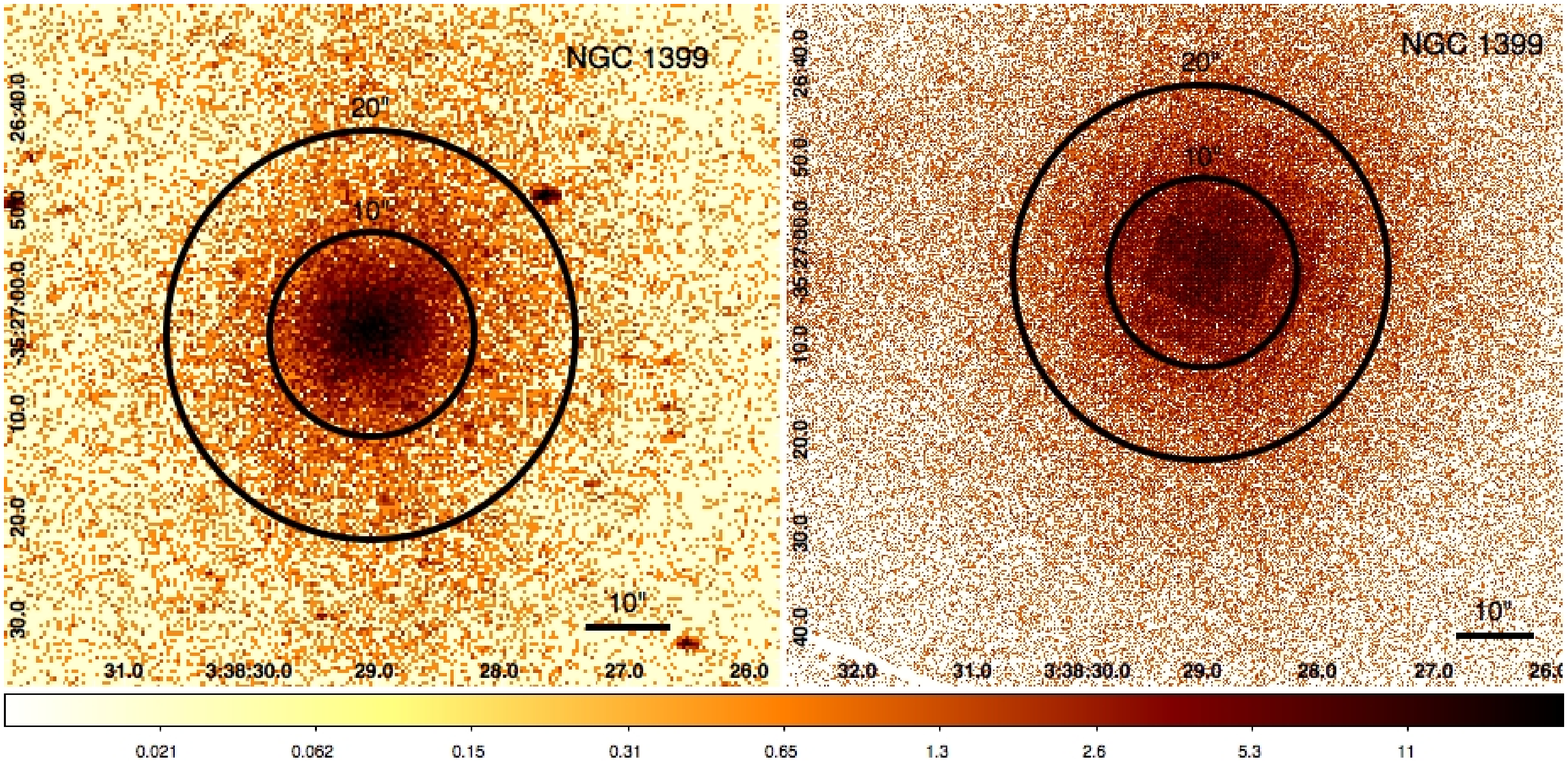}
\includegraphics[scale=0.25]{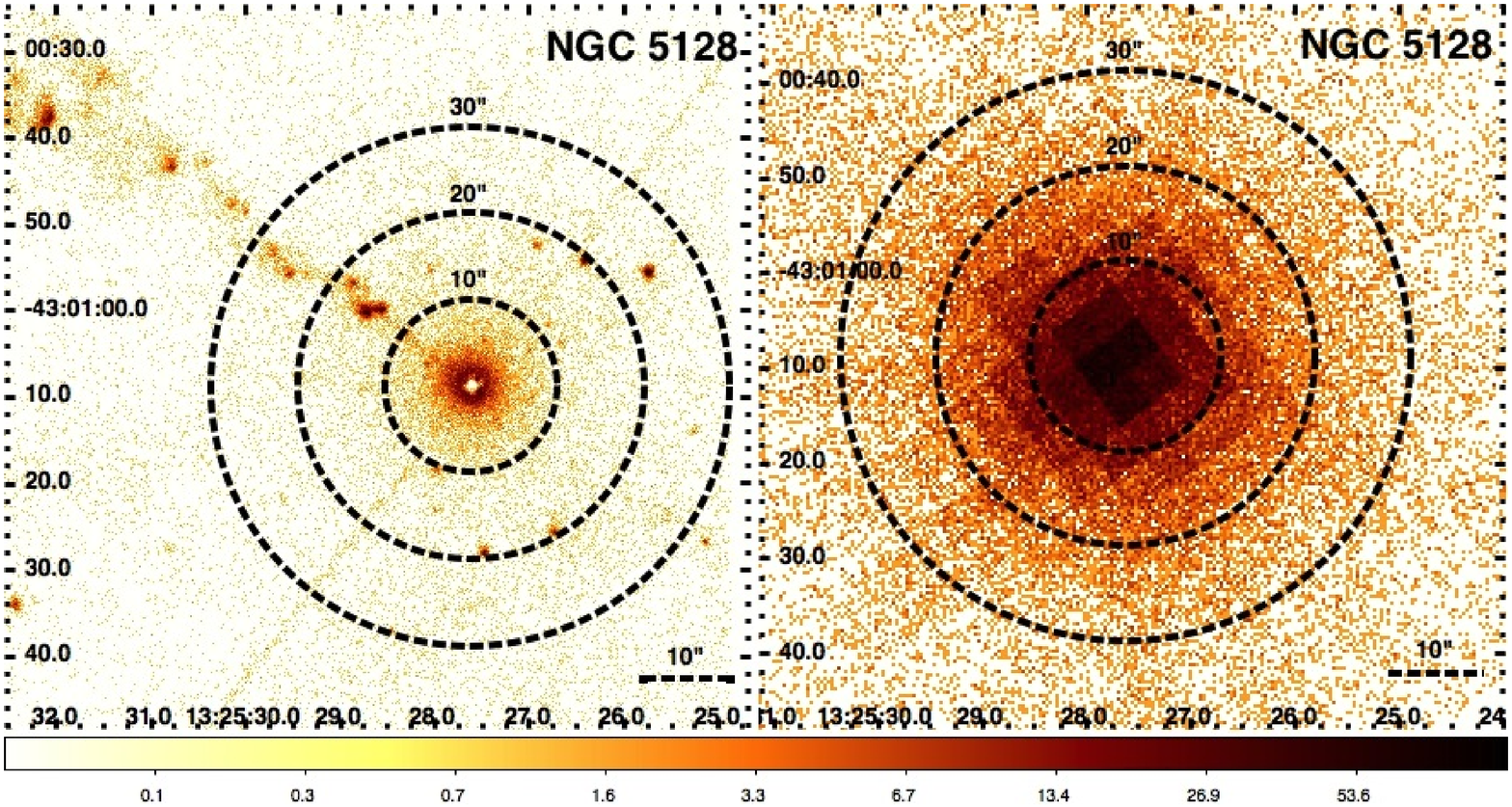}
\end{center}
\caption{Images of NGC 4151, NGC 221, NGC 1399, and NGC 5128. 
The left hand panels are \chandra\ images and the right hand panels 
\xmm\ images. 
Circles with different lines and colors indicate the extended regions for 
count estimations and nearby sources for the NGC 221 case. 
NGC 4151 is an example of a clear nuclear source, NGC 221 of a LLAGN 
surrounded by nearby objects, NGC 1399 of an AGN with substantial 
extended emission, and NGC 5128 of a 
LLAGN with jet-like emission presences.}
\label{FIG1}
\end{figure*} 

We illustrate this procedure 
in Figure \ref{FIG1} 
using NGC 4151 as an example of a clearly isolated nuclear source, 
NGC 221 (also known as M32) as an example of a LLAGN surrounded 
by bright nearby objects, NGC 1399 as an example of diffuse 
emission, and NGC 5128 showing a LLAGN with a jet-like structure. 
\chandra\ and \xmm\ images are in the left and right hand panels, respectively. 

 NGC 4151 is a well-known nearby \citep[$D\sim13.3$ Mpc,][]{mun99} 
Seyfert 1.5 galaxy \citep{OK76}. 
Both images show that it is an isolated source. 
The \chandra\ net count difference of $\sim20\%$ between the 
extraction region with radius of $1.5"$ ($1.3\times10^5$) and 
that with radius of $20"$ ($\sim1.7\times10^5$) 
confirms that in NGC 4151 the contribution from the extended
emission is negligible. Out of 53 LLAGNs, 22 sources appear
to have an isolated central source. 

The \chandra\ image of NGC 221 (top right panels of Figure 
\ref{FIG1}) shows the presence of two additional sources 
inside of the $20"$ radius. They are indicated by solid lined circles.
A source brighter than the AGN in NGC 221, `source1' in Figure 
\ref{FIG1}, is located $8.3"$ southeast of NGC 221 
and a dimmer source (`source2') $20"$ away in the western direction. 
Although source1 and the nucleus of NGC 221 are clearly 
distinguishable with the sub-arcsecond spatial resolution of \chandra, 
the \xmm\ image on the right hand panel reveals 
a single emission component. In this case, the \xmm\ spectrum is 
dominated by the brightest off-nuclear source. 
As a consequence, the \xmm\ observation of NGC 221 
cannot be used to characterize the properties of the LLAGN. 
Using \chandra\ data, we find that $92\%$ of total counts 
($1.4\times10^4$) in $10"$ are from source 1 
whereas the counts of NGC 221 are $6.1\times10^2$. 
There were a total of 23 objects containing off-nuclear sources 
within $10"-20"$ radii from the central source. 
For six objects (NGC 2787, NGC 4278, 
NGC 4374, NGC 4945, NGC 4649, and NGC 5576) the central 
source emission dominates and the contribution from off-nuclear 
sources appears to be negligible. Therefore, in these cases, the 
\xmm\ data can be used for the spectral analysis. 
For the remaining 5 sources (NGC 221, NGC 224, NGC 1023, 
NGC 3585, NGC 4291) only \chandra\ data can be used to 
investigate the AGN spectral X-ray properties given the 
significant contamination in the \xmm\ data.

The left and right bottom panels of Figure \ref{FIG1} show the 
cases of two sources (NGC 1399 and NGC 5128) whose X-ray 
emission in the \xmm\ extraction region is severely contaminated 
by extended emission and jet emission, respectively. Despite the 
fact that 12 sources are also classified as radio galaxies, 
only 3 (NGC 4486, NGC 4594, NGC 5128) show the presence of 
an extended jet-like structure in the \chandra\ images. 

In summary, after the visual inspection of all \chandra\ images of our 
sample, 22 appear to be isolated sources (indicated by ``iso" 
in the X-ray morphology classification reported in 
Table \ref{tab2}), 23 LLAGNs contain off-nuclear sources within 20" 
(``off" in Table \ref{tab2}), and the remaining have either significant 
extended emission (``ext" in Table \ref{tab2}) or jet-like 
structures (``jet"). 

\section[]{X-ray spectral analysis}
We extracted spectra in the energy range from $0.3-10$ keV 
for all isolated sources and for those without significant off-nuclear 
contribution with \xmm\ data (24 sources), and for the remaining 
29 sources we used \chandra\ data. 
We grouped 20 or 15 counts per bin, 
which is appropriate for the use of the $\chi^2$ statistics. 
To increase the statistics for the \xmm\ observation, 
we fitted simultaneously the EPIC pn, MOS1, and MOS2 spectra.
For \chandra\ spectra if the number of counts per extraction region 
was low (e.g., $\le100$), the spectra were kept ungrouped and 
the C-stat statistic was used \citep{ca79}. 
Overall, we used the C-statistics for 24 \chandra\ spectra and 
2 \xmm\ spectra. 
All sources in our sample were systemically 
fitted with a base-line model comprising a power law (PL) and 
two absorption models, 
one fixed at the Galactic value and 
the other left free to vary to mimic the intrinsic local absorption. 
When necessary, Gaussian components were 
added to fit line-like features.

The spectral results are reported 
in Table \ref{tab2}. For sources with unconstrained intrinsic 
absorption value, we reported the upper limits. 
The vast majority of the sample have X-ray photon 
indices in the range from 1 to 3, 
with a few objects yielding very hard values ($\Gamma<1$). 
Unabsorbed luminosities \lx\ are in the range of $10^{38}-10^{43}$ \ergs, 
with the exceptions of NGC 221, NGC 224, and NGC 4486A that have low 
luminosities of order of $10^{36}-10^{37}$ \ergs. Overall, the 
spectral fits of 53 sources were in the 
range of $0.8\le\chi^2_{red}\le1.5$ for the $\chi^2$ statistics 
and the C-statistic/dof also was in a similar range. 

\begin{table*}
\footnotesize
\begin{adjustwidth}{-0.8 in}{}
\caption{Spectral Analysis}
\begin{center}
\footnotesize
\begin{tabular}{lcrccclccc} 
\hline        
\hline
\noalign{\smallskip}
Source Name & Instrument & OSB ID & $N_{\rm H}$ & $\Gamma$ & $\log(L_{\rm x})$ 
& \multicolumn{1}{c}{Statistics} & X-ray Mor. & $\log(L_{\rm X}/L_{\rm Edd})$ & $R_{\rm X}$\\
\multicolumn{1}{c}{(1)} & \multicolumn{1}{c}{(2)} &  \multicolumn{1}{c}{(3)} & 
\multicolumn{1}{c}{(4)} &(5) & (6) & \multicolumn{1}{c}{(7)} & (8) & (9) & (10)\\
\noalign{\smallskip}
\hline
IC 1459 & X & 0135980201 & $ 0.16 \pm 0.01 $ & $ 1.99 \pm 0.02 $ & 40.76 & 576.06/574  & iso & -6.79 & -1.00 \\
IC 4296 & X & 0672870101 & $ 0.09 \pm 0.01 $ & $ 1.49 \pm 0.02 $ & 41.47 & 733.20/630  & iso & -5.77 & -2.88 \\
NGC 221 & C & 5690 & $  \le 0.02 $ & $ 2.29 \pm 0.16 $ & 35.92 & 32.90/24  & off & -8.68 & -2.62 \\
NGC 224 & C & 1575 & $ 0.03 \pm 0.01 $ & $ 2.50 \pm 0.09 $ & 36.77 & 63.86/56  & off & -9.51 & -4.63 \\
NGC 821 & C & 6314 & $  \le 0.16 $ & $ 1.76 \pm 0.48 $ & 38.67 & 38.70/35 $^c$ & off & -7.07 &  \\
NGC 1023 & C & 8464 & $ 0.08 \pm 0.03 $ & $ 2.14 \pm 0.12 $ & 38.66 & 135.36/179 $^c$ & off & -7.11 &  \\
NGC 1068 & C & 344 & $  \le  0.001 $ & $ 2.56 \pm 0.02 $ & 40.77 & 305.01/210  & ext & -4.27 & -1.59 \\
NGC 1300 & C & 11775 & $ 3.83 \le 0.77 $ & $ 2.24 \pm 0.14 $ & 39.68 & 85.37/107 $^c$ & iso & -6.28 &  \\
NGC 1399 & X & 0400620101 & $ 0.17 \pm 0.01 $ & $ 3.02 \pm 0.05 $ & 39.87 & 1000.58/753  & iso & -6.95 &  \\
NGC 2748 & C & 11776 & $ 0.15 \pm 0.14 $ & $ 2.34 \pm 0.65 $ & 38.30 & 11.75/17 $^c$ & off & -7.48 &  \\
NGC 2778 & C & 11777 & $  \le 0.20 $ & $ 2.29 \pm 0.57 $ & 38.43 & 30.36/28 $^c$ & off & -6.89 &  \\
NGC 2787 & X & 0200250101 & $ 0.06 \pm 0.02 $ & $ 2.03 \pm 0.06 $ & 39.09 & 131.82/115  & off & -6.66 & -1.87 \\
NGC 3031 & X & 0657801801 & $ 0.05 \pm 0.01 $ & $ 2.06 \pm 0.01 $ & 40.27 & 999.93/896  & iso & -5.74 & -3.45 \\
NGC 3115 & C & 11268 & $ 0.03 \pm 0.03 $ & $ 2.11 \pm 0.20 $ & 38.35 & 97.60/125 $^c$ & off & -8.74 &  \\
NGC 3227 & X & 0400270101 & $ \cdots   $ & $ 1.41 \pm 0.003 $ & 42.06 & 2050.46/1890  & iso & -3.23 & -4.34 \\
NGC 3245 & C & 2926 & $  \le 0.23 $ & $ 1.83 \pm 0.36 $ & 39.27 & 39.93/57 $^c$ & iso & -7.19 & -2.29 \\
NGC 3377 & C & 2934 & $ 0.15 \pm 0.05 $ & $ 2.06 \pm 0.23 $ & 38.29 & 66.62/93 $^c$ & off & -7.88 &  \\
NGC 3379 & C & 7076 & $ 0.10 \pm 0.04 $ & $ 2.26 \pm 0.21 $ & 38.33 & 89.83/113 $^c$ & off & -7.87 &  \\
NGC 3384 & C & 11782 & $  \le 0.21 $ & $ 1.83 \pm 0.27 $ & 38.55 & 51.15/73 $^c$ & off & -6.81 &  \\
NGC 3585 & C & 9506 & $ 0.08 \pm 0.03 $ & $ 2.39 \pm 0.20 $ & 38.78 & 103.00/115 $^c$ & off & -7.86 &  \\
NGC 3607 & X & 0099030101 & $ 0.06 \pm 0.04 $ & $ 2.63 \pm 0.22 $ & 38.85 & 33.15/35  & iso & -7.34 &  \\
NGC 3608 & C & 2073 & $  \le  0.23 $ & $ 2.44 \pm 0.31 $ & 38.58 & 42.98/63 $^c$ & ext & -7.85 &  \\
NGC 3998 & X & 0090020101 & $ 0.01 \pm 0.01 $ & $ 1.85 \pm 0.01 $ & 41.43 & 1159.15/1163  & iso & -5.05 & -3.40 \\
NGC 4026 & C & 6782 & $  \le 0.33 $ & $ 2.11 \pm 0.55 $ & 38.06 & 21.05/29 $^c$ & ext & -8.38 &  \\
NGC 4151 & X & 0143500301 & $  \le 0.08 $ & $ -0.46 \pm 0.00 $ & 42.79 & 1899.16/1594  & iso & -2.97 & -4.59 \\
NGC 4258 & X & 0400560301 & $ 0.38 \pm 0.01 $ & $ 2.10 \pm 0.34 $ & 40.45 & 1253.17/1142  & iso & -5.24 & -4.42 \\
NGC 4261 & X & 0056340101 & $  \le 0.02 $ & $ 0.82 \pm 0.03 $ & 40.94 & 561.75/441  & iso & -5.91 & -1.73 \\
NGC 4278 & X & 0205010101 & $ 0.02 \pm 0.01 $ & $ 2.06 \pm 0.01 $ & 40.26 & 934.64/929  & off & -7.05 & -2.35 \\
NGC 4291 & C & 11778 & $  \le 0.10 $ & $ 2.11 \pm 0.23 $ & 39.12 & 75.50/87 $^c$ & off & -7.50 &  \\
NGC 4303 & X & 0205360101 & $  \le 0.18 $ & $ 2.90 \pm 0.37 $ & 39.06 & 49.45/43  & iso & -5.70 & -0.60 \\
NGC 4342 & C & 12955 & $  \le 0.08 $ & $ 1.90 \pm 0.15 $ & 38.62 & 159.76/182 $^c$ & off & -8.05 &  \\
NGC 4374 & X & 0673310101 & $  \le 0.02 $ & $ 2.16 \pm 0.05 $ & 39.68 & 347.37/315  & off & -7.61 & -0.91 \\
NGC 4395 & X & 0142830101 & $ 0.24 \pm 0.01 $ & $ 0.96 \pm 0.01 $ & 40.16 & 3030.09/2394  & iso & -2.99 & -4.60 \\
NGC 4459 & X & 0550540101 & $ 0.20 \pm 0.02 $ & $ 1.99 \pm 0.09 $ & 39.37 & 147.63/162  & iso & -6.61 &  \\
NGC 4473 & C & 4688 & $  \le  0.28 $ & $ 1.70 \pm 0.52 $ & 38.97 & 22.96/28 $^c$ & ext & -7.25 &  \\
NGC 4486 & C & 2707 & $  \le 0.001 $ & $ 1.67 \pm 0.01 $ & 41.24 & 582.59/400  & jet & -6.43 & -1.41 \\
NGC 4486A & C & 11783 & $  \le 0.21 $ & $ 1.47 \pm 0.29 $ & 37.00 & 57.15/61 $^c$ & ext & -8.24 &  \\
NGC 4564 & C & 4008 & $  \le 0.15 $ & $ 1.75 \pm 0.29 $ & 38.89 & 64.17/61 $^c$ & off & -7.06 &  \\
NGC 4594 & X & 0084030101 & $ 0.14 \pm 0.01 $ & $ 1.85 \pm 0.02 $ & 40.51 & 551.35/574  & jet & -6.36 & -2.62 \\
NGC 4596 & C & 11785 & $  \le 0.16 $ & $ 1.48 \pm 0.24 $ & 39.28 & 58.83/69 $^c$ & off & -6.75 &  \\
NGC 4649 & X & 0502160101 & $ 0.06 \pm 0.02 $ & $ 2.55 \pm 0.02 $ & 40.06 & 1254.12/811  & off & -7.38 & -2.61 \\
NGC 4697 & C & 4730 & $  \le 0.16 $ & $ 1.41 \pm 0.19 $ & 39.00 & 68.99/94 $^c$ & off & -7.40 &  \\
NGC 4742 & C & 11779 & $ 0.04 \pm 0.04 $ & $ 1.76 \pm 0.17 $ & 39.15 & 95.84/135 $^c$ & off & -6.14 &  \\
NGC 4945 & X & 0204870101 & $  \le 0.03 $ & $ 0.03 \pm 0.03 $ & 40.13 & 633.61/502  & off & -4.13 & -1.96 \\
NGC 5077 & C & 11780 & $ 0.14 \pm 0.06 $ & $ 1.88 \pm 0.23 $ & 39.73 & 78.05/83 $^c$ & iso & -7.28 &  \\
NGC 5128 & C & 3965 & $ 5.50 \pm 0.13 $ & $ -0.27 \pm 0.02 $ & 40.89 & 183.32/216  & jet & -5.70 & -1.04 \\
NGC 5252 & X & 0152940101 & $ 2.08 \pm 0.01 $ & $ 1.11 \pm 0.01 $ & 43.11 & 2226.37/2050  & iso & -4.00 & -4.06 \\
NGC 5576 & X & 0502480701 & $ \cdots   $ & $ 1.83 \pm 0.19 $ & 39.21 & 295.30/325 $^c$ & off & -7.16 &  \\
NGC 5845 & X & 0021540501 & $  \le 0.07 $ & $ 2.27 \pm 0.34 $ & 38.85 & 163.04/209 $^c$ & iso & -7.72 &  \\
NGC 6251 & X & 0056340201 & $ 0.04 \pm 0.01 $ & $ 1.93 \pm 0.02 $ & 42.67 & 1211.77/1069  & iso & -4.22 & -1.66 \\
NGC 7052 & C & 2931 & $ 0.06 \pm 0.04 $ & $ 2.71 \pm 0.27 $ & 40.24 & 68.57/79 $^c$ & iso & -6.47 & -0.81 \\
NGC 7457 & C & 11786 & $  \le 0.19 $ & $ 1.41 \pm 0.33 $ & 38.36 & 25.89/39 $^c$ & iso & -6.36 &  \\
NGC 7582 & X & 0204610101 & $ 0.12 \pm 0.01 $ & $ 0.07 \pm 0.03 $ & 41.13 & 1581.04/1044  & iso & -4.72 & -2.58 \\
\hline
\end{tabular}
\end{center}
\footnotesize
\label{tab2}
\begin{itemize}
\item[]
\begin{itemize}
\textbf{Columns} 1 = source name, 2= observation ID, 3 = counts in source region,  
4= intrinsic absorption value in units of $10^{22}$ cm$^{-2}$, 5 = photon index, 
6 = unabsorbed luminosity in $2-10$ keV, 
7 = $\chi^2$/degree of freedom: $^c$ C-statistic/degree of freedom, 
8 = X-ray morphology (iso $-$ central AGN; 
off $-$ presence of nearby objects within 10 arc sec; 
ext $-$ diffused emission, jet $-$ jet-like structure emission), 
9 = $\log(L_{\rm X}/L_{\rm Edd})$  
10 = X-ray radio-loudness parameter, $R_{\rm X}$($=\log(L_{\rm R}/L_{\rm X})$)
\end{itemize}
\end{itemize}
\end{adjustwidth}
\end{table*}

For completeness, for sources with both \chandra\ and \xmm\ 
observations, we have compared the photon index and X-ray flux 
measured by the two satellites at different epochs. The vast majority 
of the AGNs shows consistent values with variations within the $3\sigma$ 
level. The only discrepancies are observed in objects that have poorly 
constrained \chandra\ spectra with \gam\ either very steep ($>3$) or 
very flat ($<1$ or inverted spectra). We can therefore conclude that 
flux and spectral variability does not significantly affect our analysis.

\begin{figure}
\includegraphics[scale=0.5]{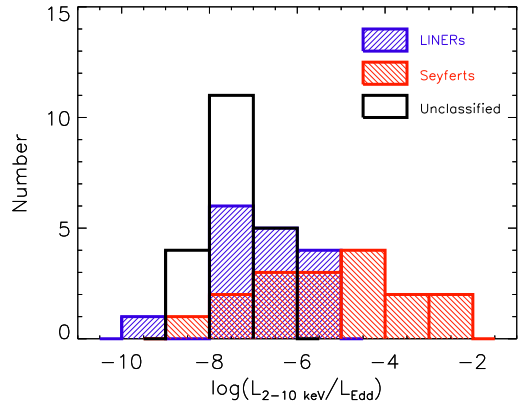}
\caption{Distribution of \edd. The histogram filled with 
positive slopes indicate LINERs, the negative slopes filled 
represent Seyferts, and the empty one unclassified AGNs.}
\label{FIG2}
\end{figure}

Combining the X-ray luminosities inferred from the spectral analysis
with the black hole masses from dynamical measurements, we derive 
the X-ray Eddington ratio, \edd\ ($L_{\rm 2-10\, keV}/L_{\rm Edd}$), 
for all the sources. Throughout the paper we use \lx\ to indicate the 
luminosity in the $2-10$ keV energy range, which is the most common 
band used in X-ray studies and allow a direct comparison with literature results.
These values of $\log(L_{\rm X}/L_{\rm Edd})$, reported in Table \ref{tab2}, 
range between $-10$ and $-3$, with a mean of $-6.5\pm1.5$. 
The distribution of $\log(L_{\rm X}/L_{\rm Edd})$ for each optical class 
is plotted in Figure \ref{FIG2}. 
Assuming a bolometric correction of $15-30$, which is appropriate for 
low-accreting AGNs (see, e.g., Vasudevan \& Fabian 2009), we obtain 
Eddington ratio values $\log(L_{\rm bol}/L_{\rm Edd})$ ranging between  
$-8.8$ and $-1.8$ with a mean of $-5.3$, which confirms
that this sample comprises only low-accreting AGNs.

Very flat X-ray spectra are often associated with heavily 
absorbed AGNs. In the most extreme cases (i.e., for 
Compton-thick sources with $N_{\rm H}>10^{24}\,{\rm cm}^{-2}$), 
the direct coronal emission is completely absorbed and the detected 
X-rays are thought to be produced by reflection. Since 
in these sources the estimated \lx\ is severely underestimated 
and \gam\ is not representative of the direct emission, it is not 
possible to extend the X-ray scaling method to 
Compton-thick AGNs. For this reason, it is crucial to identify 
(and exclude from further analysis) 
Compton-thick sources. Typically, two different approaches are used 
to find Compton-thick candidates: 1) the detection of Fe K$\alpha$ 
lines with large equivalent width (EW$>1$keV) and 2) the use of the 
$T_{\rm ratio}=F_{\rm 2-10 keV}/F_{\rm [OIII]}$ parameter 
(where $F_{\rm [OIII]}$ is corrected for optical reddening), 
with the assumption that the X-ray flux is associated with the 
absorbed AGN component, whereas the O[III] flux is considered 
a reliable indicator of the isotropic emission since it is mostly 
produced in the unobscured narrow line region (Bassani et al. 1999). 
Past studies have shown that Compton-thick objects are characterized 
by values of T below 1. 

\begin{figure}
\includegraphics[scale=0.5]{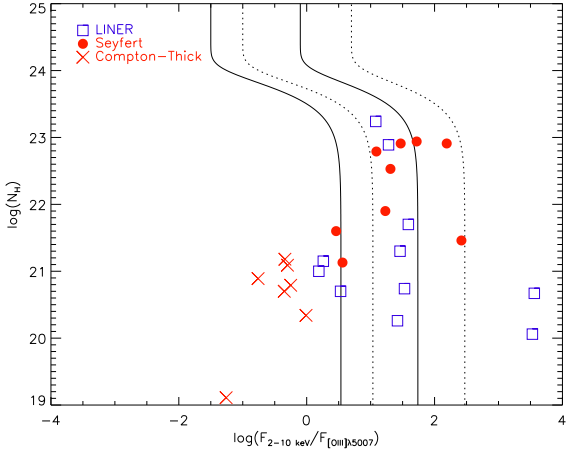}
\caption{Plot of $\log(N_{\rm H})$ vs. $\log(F_{\rm 2-10\,keV}/F_{\rm [OIII]})$. 
The filled circles indicate Seyfert galaxies and the open squares LINERs. 
The cross mark ``x" was used to indicate the Compton-thick candidates. 
The solid lines indicate the expected correlation derived by Cappi et al. (2006) 
assuming that \lx\ is absorbed by the measured $N_{\rm H}$ and a $1\%$ 
reflected component. Similarly the dashed lines indicate the correlation derived 
by Maiolino et al. (1998).}
\label{fig3}
\end{figure} 

We have computed the T factor for all the 
objects of our sample with optical line measurements. The results 
are plotted in Figure \ref{fig3} where the lines 
represent the expected correlation between T and $N_{\rm H}$ 
for Seyfert galaxies under the assumption that the X-ray flux is 
absorbed by the measured $N_{\rm H}$. Figure \ref{fig3}, combined 
with results from the spectral analysis showing flat 
spectra and in some cases a Fe K$\alpha$ line with large EW, 
suggests that 8 sources (NGC 1068, NGC 2748, NGC 3607, 
NGC 3245, NGC 4303, NGC 4374, NGC 4945, NGC 7582) 
may be genuine Compton-thick candidates, in agreement with 
independent findings in the literature (Bianchi et al. 2009; 
Cappi et al. 2006; Gon$\acute{\rm z}$alez-Martin et al. 2009; 
Marinucci et al. 2012; Yaqoob 2012). 
To be conservative, we exclude from
further analysis these 8 objects.

Sources with flat spectra but without evidence for Compton thickness
and sources that showed substantial residuals when fitted with 
our simple base-line model were re-fitted with more complex models. 
These spectral models may comprise a thermal component (apec 
in Xspec) to account for galaxy contribution, a blackbody to mimic 
a soft excess, a partial covering model (zpcfabs) to account 
for absorbers with patchy geometry and a reflection component (pexrav). 
This additional spectral analysis 
yielded steeper photon indices as indicated by Table \ref{tab3} 
that reports the most relevant spectral parameters.

In summary, all sources were reasonably well fitted by either an absorbed
power law or slightly more complex models, yielding \gam\ values in the
range $1.3-3$ and \lx\ between $10^{37}$ and $10^{43}$ \ergs. 

\begin{table*}
\begin{adjustwidth}{-0.8 in}{}
\footnotesize
\caption{Spectral Analysis $-$ For Objects with Very Flat Spectra}
\begin{center}
\footnotesize
\begin{tabular}{lccccccccl}
\hline        
\hline
\noalign{\smallskip}
Source Name & Model & $N_{\rm H}$ & R & E$_{\rm break}$ & \gam & FeK$\alpha$ & EW & 
$\log(L_{\rm x})$ & \multicolumn{1}{c}{$\chi^2$/dof}\\
\multicolumn{1}{c}{(1)} & (2) &  (3) & (4) &(5) & (6) & (7) & (8) &(9) 
& \multicolumn{1}{c}{(10)}\\
\noalign{\smallskip}
\hline
NGC 5128 & wabs(bknpo+pexrav) & $8.09\pm0.27$ & $0.63\pm0.09$& $4.68\pm0.07$ 
& $0.22\pm0.01$ & $6.42\pm0.03$& $0.03$&  40.88 &  191.36/200\\
\hline
\hline
Source Name & Model & $N_{\rm H}$ & CF & kT & 
\gam & Fe K$\alpha$ & EW & $\log(L_{\rm x})$ & \multicolumn{1}{c}{$\chi^2$/dof}\\
&  & & (11) & (12) & &\\
\hline
NGC 1300 & pcfabs(pow) & $3.83\pm0.77$ & $0.96\pm0.02$ &  & $2.24\pm0.14$ &  &  & 39.68 & 85.37/107$^c$\\
NGC 3031 & wabs(bb+pow) & $0.00$ &  & $0.29\pm0.02$ & $1.90\pm0.01$ &  &  & 39.38 & 985.49/895\\
NGC 3115 & pcfabs(pow) & $0.13\pm0.08$ & $0.70\pm0.33$ &  & $2.34\pm0.24$ &  &  & 38.09 & 80.94/126\\
NGC 3227 & pcfabs(pow)+gauss & $6.16\pm0.15$ & $0.91$ &  & $1.39\pm0.01$ & $6.38\pm0.01$ & $0.06\pm0.02$ & 41.77 & 1866.78/1739\\
NGC 4151 & pcfabs(pow+apec+gauss) & $3.38\pm0.02$ & $0.95$ & $0.14$ & $1.33\pm0.01$& $6.37\pm0.01$ & $0.06\pm0.02$ & 42.88 & 2617.68/1892\\
NGC 4258 & pow+wabs(pow) & $8.74\pm0.20$ &  &  & $1.70\pm0.02$ &  &  & 40.51 & 966.56/945\\
NGC 4261 & pow+pcfabs(pow) & $7.71\pm0.62$ &  &  & $1.65\pm0.03$ &  &  & 41.16 & 457.95/442\\
NGC 4342 & pcfabs(pow) &  & $0.78\pm0.02$ &  & $1.61\pm0.12$ &  &  & 39.81 & 172.65/173$^c$\\
NGC 4395 & pcfabs(apec+pow) & $0.79\pm0.02$ & $0.72$ & $0.19$ & $1.11\pm0.01$ &  &  & 39.96 & 3197.11/2394\\
NGC 4596 & pcfabs(pow) & $17.20$ & $0.05$ &  & $1.89\pm0.36$ &  &  & 38.86 & 36.87/67\\
NGC 4649 & pcfabs*apec*pow & $0.13\pm0.01$ & $0.82\pm0.02$ & $0.87$ & $1.64\pm0.03$ &  &  & 40.12 & 1250.62/798\\
NGC 4697 & pcfabs(pow) & $0.19\pm0.07$ & $\le0.79$ &  & $2.08\pm0.30$ &  &  & 38.25 & 50.27/91\\
NGC 5128 & abs(apec+pow) & $8.14\pm0.16$ &  & $0.11$ & $0.32\pm0.01$ &  &  & 40.40 & 245.92/216\\
NGC 6251 & pcfabs*apec*pow & $0.40\pm0.03$ & $0.54\pm0.02$ & $0.61\pm0.03$ & $1.83\pm0.01$ &  &  & 42.72 & 1143.05/1059\\
NGC 7052 & pcfabs(bb+pow) & &  & $0.16\pm0.03$ & $1.76\pm0.36$ &  &  & 40.43 & 13.46/15\\
NGC 7457 & wabs(apec+pow) & $0.29\pm0.09$ &  & $0.13\pm0.02$ & $1.52\pm0.39$ &  &  & 38.50 & 11.17/12\\
\hline
\hline
\end{tabular}
\end{center}
\footnotesize
\label{tab3}
\begin{itemize}
\item[]
\textbf{Columns} 1 = source name, 2 = model used, 
3 = intrinsic absorption value in units of $10^{22}$ cm$^{-2}$, 
4 = Reflection Factor ($0\le{\rm R}\le1$: 0 $-$ no reflection component and 1 $-$ 
isotropic source above disk),
5 = energy break (keV),
6 = photon index,   7 = Fe K$\alpha$ line (keV), 
8 = equivalent width (keV), 9 = unabsorbed luminosity in $2-10$ keV, 
10 = $\chi^2$/degree of freedom ($^c$ indicates C-statistic/degree of freedom),
11 = demensionless coverage fraction, 12 = temperature of soft excess in units of keV.\\
\end{itemize}
\end{adjustwidth}
\end{table*}

\subsection{X-ray Radio-loudness}
Combining the X-ray luminosity inferred from the spectral analysis 
and the radio luminosity from the literature, we computed the X-ray 
radio loudness parameter, $R_{\rm X}=L(6 \rm {cm})/L_{2-10\,{\rm keV}}$, which 
was introduced by Terashima \& Wilson (2003) to reduce the extinction 
that affects the optical emission used in the classical radio-loudness parameter. 
The values of $R_{\rm X}$ for each objects are reported in Table \ref{tab2}.
For the 26 sources (9 LINERs, 14 Seyferts, 3 unclassified) for which the radio data
are available, $\log(R_{\rm X})$ ranges between $-5$ and $0$ with a mean of 
$-2.5\pm1.3$. These values are consistent with those of the sample of low-luminosity 
Seyfert galaxies analyzed by Panessa et al. (2007). Similar to Panessa et al. (2007) 
we did not find any correlation between $R_{\rm X}$ and \lx, whereas there is suggestive 
evidence for an anti-correlation between $R_{\rm X}$ and \edd. However, unlike Panessa 
et al. (2007) the anti-correlation in our sample is not statistically significant: the negative 
slope is consistent with 0 within $2\sigma$s. This can be explained either 
but the limited number of objects with radio data in our sample or by the fact that our objects 
are accreting at a much lower level and in this regime no correlation is expected between 
radio-loudness and Eddington ratio, as demonstrated by Sikora et al. (2007).

\begin{figure} 
\includegraphics[scale=0.5]{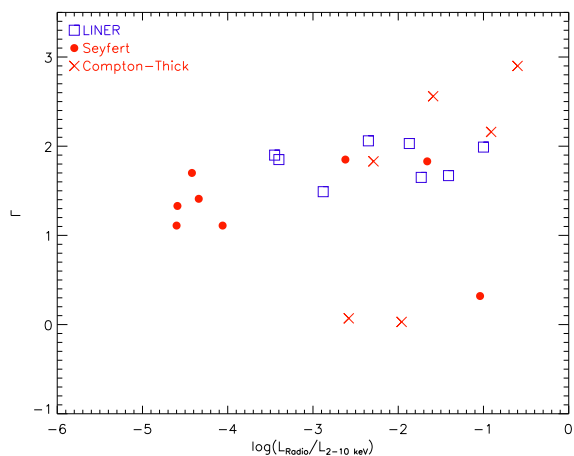}
\caption{Plot of $\Gamma$ vs. $\log(L(6 \rm {cm})/L_{2-10\,{\rm keV}})$. 
Open squares were used to indicate LINERs, filled circles for Seyferts, and 
crosses for Compton-thick candidates.}
\label{Radiolouness}
\end{figure}

We also looked for any correlation between \gam\ and $R_{\rm X}$ 
and plotted of $\Gamma$ versus $R_{\rm X}$ is in Figure 
\ref{Radiolouness}. When all objects with $R_{\rm X}$ are included, there is no evidence 
for any correlation (the Spearman's $\rho-$rank is $0.45$ with a probability of $P=0.03$). 
If we exclude the Compton-thick candidates and the source 
with very low \gam, then a weak positive correlation of $\Gamma=(0.17\pm0.05)R_{\rm X}+(2.16\pm0.22)$ 
with the RMS value of $0.22$ is found.

\section{Estimation of \mbh\ with X-rays}
\subsection{X-ray Scaling Method}
In a recent study on GBHs, \cite{shapo09} showed that 
spectral transitions of different GBHs present two similar 
positive correlations between temporal and spectral properties: 
1) a correlation between the quasi periodic oscillation (QPO) 
frequency and the photon index and 2) a correlation between 
$N_{BMC}$, the normalization of the bulk \comp\ (BMC) model, 
and the photon index. Because different BHs show similar trends 
in the $\Gamma-\mathrm{QPO}$ and $\Gamma-N_{BNC}$ diagrams, 
it can be shown that $M_{BH}$ (and distance) of any GBH can 
be determined by simply shifting this self-similar function 
until it matches the spectral pattern of a GBH of known 
$M_{BH}$ and distance (considered as a reference source).

In the following, we briefly describe the \comp\ model used 
in the X-ray scaling method, the basic characteristics of the 
technique utilized to estimate \mbh, and the main results 
obtained applying this method to AGNs. 

It is widely accepted that the X-ray emission associated with 
BH systems is produced by the \comp\ process in the corona. 
The BMC model is a simple and robust \comp\ model which equally 
well describes thermal and bulk \comp\ \citep{titar97}. 
The BMC model is characterized by four parameters: 
(1) the temperature of the thermal seed photons, $kT$, 
(2) the energy spectral index $\alpha$ (related to the photon 
index by the relation \gam$=\alpha+1$), 
(3) $\log(A)$ which is related to the \comp\ fraction by $f=A/(1+A)$ 
(where $f$ is the ratio between the number of Compton scattered 
photons and the number of seed photons), and 
(4) the normalization, \nbmc\ which depends on the luminosity and 
the distance, (\nbmc\ $\propto L/d^2$). 

The X-ray scaling method, based on two diagrams \gam$-\nu_{QPO}$ 
and \gam$-$\nbmc, allowed the estimate of \mbh\ and distance in 
several GBHs by scaling the temporal  and spectral properties of a 
reference GBH. However, the much longer timescales of AGNs and 
the absence of detectable QPOs do not allow the use of the \gam\ 
$-$ QPO diagram. On the other hand, the \gam\ $-$ \nbmc\ plot 
can be easily extended to AGNs assuming that AGNs follow a 
similar spectral evolution as GBHs. This method relies on the 
direct dependence of the BMC normalization on \mbh:
$N_{BMC}=L/d^2$, where $L\propto\eta\dot{m}M_{BH}$ is the 
accreting luminosity, $\eta$ is the radiative efficiency and $\dot{m}$ 
the accretion rate in Eddington units. Therefore, the normalization, 
\nbmc, can be expressed as a function of $M_{BH}$: 
$N_{BMC}\propto\eta\dot{m}M_{BH}/d^2=M_{\rm BH}/d^2$, 
which is derived assuming that different BHs in the same spectral 
state (defined by the value of \gam) are characterized by similar 
values of $\eta$ and $\dot{m}$. 

With this X-ray scaling method, \cite{glio11} estimated $M_{BH}$ 
for a well defined sample of AGNs accreting at moderate/high 
level ($\dot{m}\gg 1\%$) whose BH mass was determined via 
reverberation mapping and which had good quality \xmm\ archival data.
The good agreement between the $M_{BH}$ values determined 
by these two methods (the RMS around the one-to-one correlation 
between $\log M_{\rm BH,scaling}$ and $\log M_{\rm BH,RM}$ 
is 0.35 using GRO J1655-40 as a reference) 
confirmed the validity of this novel technique that can 
be successfully used for both sMBHs and SMBHs.

At very low accreting rate however, X-rays are likely to be produced by 
different mechanisms (for example by advection dominated 
accretion flows, ADAFs, or can be directly related to 
jet emission). Moreover, the applicability of the X-ray scaling 
method is questionable, since it is based on the similarity of the 
spectral transition of BHs at relatively high accretion rate 
(\edd$\ge10^{-2}$), which is described by 
a positive correlation in the \gam$-$flux diagram. 

\begin{figure}
\begin{center}
\includegraphics[scale=0.5]{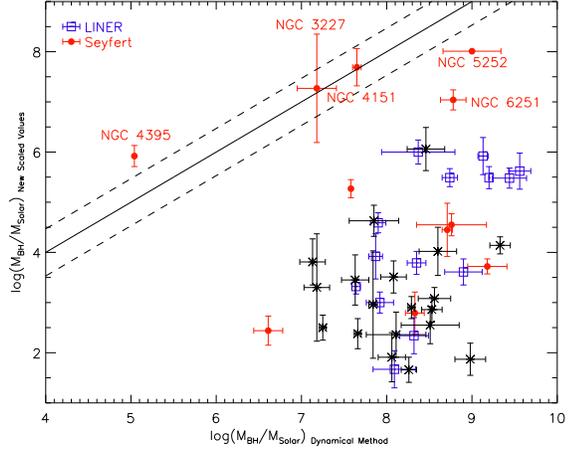}
\end{center}
\caption{\mbh\ values obtained with the scaling method plotted 
vs. \mbh\ values measured by dynamical method. 
The Seyfert sample is plotted with filled circles, open 
squares for LINERs, and crosses for unclassified LLAGNs. The 
solid line is to indicate the one-to-one correlation. Five objects 
with \edd\,$^<_{\sim}10^{-4}$ are labeled next to the data points.}
\label{FIG4}
\end{figure}

To test whether the scaling method can be applied to LLAGNs, 
we fitted their spectra with the BMC model. 
LLAGNs with \gam$=1.3-3$ were compared to 
different spectral evolution trends of GBHs.
The X-ray scaling method results are shown in Figure \ref{FIG4}. 
There is a clear 
inconsistency between the values 
from the scaling method and those obtained from dynamical methods.

The vast majority of the \mbh\ inferred from the scaling methods 
lie well below the one-to-one correlation (indicated by the solid line), 
and appear to be underestimated by $2-4$ orders of magnitude. 
This indicates that the X-ray scaling method cannot be used to 
constrain the \mbh\ of very low-accreting AGNs. The only 
noticeable exceptions are NGC 3227 and NGC 4151, which appears 
to be fully consistent with their corresponding dynamical 
estimate, and NGC 4395 and NGC 5252, that are marginally 
consistent. Importantly, these sources have the highest 
\edd\ values in our sample and suggest that the scaling 
method is still valid for \edd\ of the order of $10^{-3}-10^{-4}$. 

\subsection{\GLedd\ Anti-Correlation} 
\begin{figure}
\begin{center}
\includegraphics[scale=0.5]{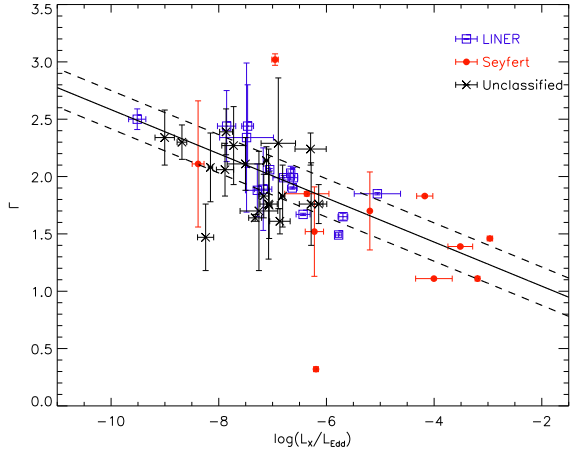}
\end{center}
\caption{Anti-correlation of $\Gamma-L_{\rm X}/L_{\rm Edd}$. 
The LINER data are indicated by open squares whereas 
the filed circles represent Seyferts and crosses 
unclassified ones. The best-fit (with parameters in Table 
for Case 2 ALL in Table \ref{tab4}) is indicated with  
the solid line with the dashed lines showing the $1\sigma$ uncertainty.}
\label{fig5}
\end{figure}

We tested whether our LLAGN sample showed any evidence 
for an anti-correlation in the \gam\ vs. \edd\ plot. 
The results, obtained using the photon index and the luminosity in the $2-10$ keV range, are 
shown in Figure \ref{fig5} and support the existence of 
an anti-correlation, whose best-fit is indicated by the solid line 
and the $1\sigma$ uncertainty with dashed lines.   

\begin{figure}
\begin{center}
\includegraphics[scale=0.5]{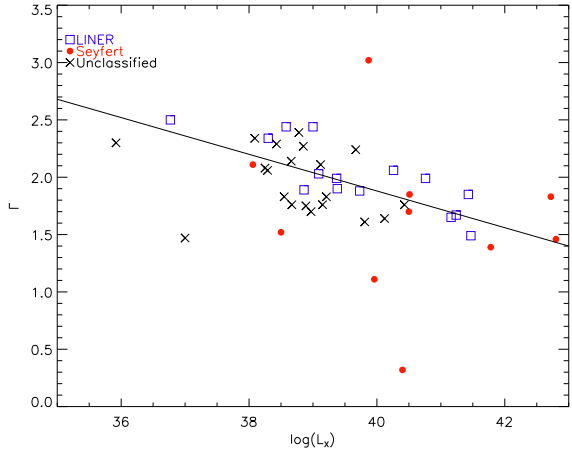}
\end{center}
\caption{Anti-correlation of $\Gamma-L_{\rm X}$. 
The LINER data are indicated by open squares whereas 
filled circles represent Seyferts and crosses for 
unclassified ones. The best-fit (with parameters 
for Case 2 ALL in Table \ref{tab4}) is indicated with  
the solid line.}
\label{GammaLx}
\end{figure}

The slope and y-intercept of best-fit results of 
LINERs, Seyferts, unclassified AGNs, and the combination of 
LINERs and Seyferts and all are reported in Table \ref{tab4}. 
There is suggestive evidence for an anti-correlation for LINERs, 
Seyferts, and the combination of all, when the results from 
the fitting of a simple PL model are used (Case 1 Table \ref{tab4}) and 
of more complex spectral models (Case 2 Table \ref{tab4}). 
When all AGN classes are combined, the significant negative 
correlations are confirmed by a non-parametric correlation 
analysis based on Spearman's 
$\rho-$rank coefficient which yields a value of 
$-0.42$ ($P=8.19\times10^{-3}$) for Case 1, and 
$-0.65$ ($P=1.44\times10^{-6}$) for Case 2.
We also tested the anti-correlation test for Case 2 without 
NGC 1399 (\gam$\approx3$) and NGC 5128 ($<1$) and the best-fit 
parameters remained consistent within the $1\sigma$ uncertainty. 
For completeness, in Table \ref{tab4} we also report the results 
of the correlation analysis between \gam\ and \lx\ (e.g., Emmanoulopoulos et al. (2012)) 
and show the $\Gamma-L_{\rm X}$ plot in Figure \ref{GammaLx}.

\begin{table}
\footnotesize
\caption{X-ray Properties Correlation Analysis Results}
\begin{center}
\footnotesize
\begin{tabular}{lccc} 
\hline        
\hline
\noalign{\smallskip}
\multicolumn{1}{c}{AGN Class} & Slope & Y-int. & RMS\\ 
\multicolumn{1}{c}{(1)}& (2) & (3) & (4)\\
\hline
\noalign{\smallskip}
\multicolumn{4}{c}{\gam$-$\edd\ Analysis Results}\\
\hline
\noalign{\smallskip}
\multicolumn{4}{c}{Case 1 $-$ use of results from the base-line model}\\
\noalign{\smallskip}
LINER & $-0.25\pm 0.09$ & $ 0.19\pm 0.62$ & 0.33\\
Seyfert & $-0.30\pm 0.16$ & $-0.17\pm 0.88$ & 0.84\\
L+S$^{\dagger}$ & $-0.29\pm 0.08$ & $-0.11\pm 0.49$ & 0.61\\
Unclassified & $-0.20\pm 0.15$ & $ 0.49\pm 1.11$ & 0.45\\
ALL & $-0.26\pm 0.06$ & $ 0.03\pm 0.39$ & 0.55\\
\noalign{\smallskip}
\hline
\noalign{\smallskip}
\multicolumn{4}{c}{Case 2 $-$ use of results from the complex model}\\
\noalign{\smallskip}
LINER & $-0.22\pm 0.05$ & $ 0.46\pm 0.32$ & 0.16\\
Seyfert & $-0.17\pm 0.11$ & $ 0.68\pm 0.63$ & 0.58\\
L+S$^{\dagger}$ & $-0.20\pm 0.05$ & $ 0.56\pm 0.34$ & 0.41\\
Unclassified & $-0.13\pm 0.08$ & $ 0.97\pm 0.60$ & 0.26\\
ALL & $-0.18\pm 0.04$ & $ 0.66\pm 0.25$ & 0.35\\
\hline
\hline
\multicolumn{4}{c}{\gam$-$\lx\ Analysis Results}\\
\noalign{\smallskip}
\hline
\multicolumn{4}{c}{Case 1}\\
\noalign{\smallskip}
LINER & $-0.21\pm 0.07$ & $10.41\pm 2.77$ & 0.32\\
Seyfert & $-0.24\pm 0.18$ & $11.24\pm 7.50$ & 0.90\\
L+S$^{\dagger}$ & $-0.26\pm 0.09$ & $12.11\pm 3.58$ & 0.66\\
Unknown & $-0.01\pm 0.12$ & $ 2.34\pm 4.74$ & 0.47\\
Comb & $-0.20\pm 0.06$ & $ 9.63\pm 2.40$ & 0.60\\
\noalign{\smallskip}
\hline
\noalign{\smallskip}
\multicolumn{4}{c}{Case 2}\\
\noalign{\smallskip}
LINER & $-0.19\pm 0.04$ & $ 9.43\pm 1.39$ & 0.17\\
Seyfert & $-0.10\pm 0.13$ & $ 5.71\pm 5.26$ & 0.62\\
L+S$^{\dagger}$ & $-0.17\pm 0.06$ & $ 8.72\pm 2.39$ & 0.45\\
Unknown & $-0.08\pm 0.06$ & $ 5.08\pm 2.47$ & 0.27\\
Comb & $-0.16\pm 0.04$ & $ 8.28\pm 1.45$ & 0.27\\
\hline
\hline
\end{tabular}
\end{center}
\footnotesize
{\bf Note:} Column List $-$ (1) AGN class; 
(2) a best-fit slope; (3) a best-fit intercept'
(4) RMS for the best-fit. All Compton-thick 
sources are excluded during the anti-correlation 
between \gam\ and \edd\ confirmation. \\
$^{\dagger}$ $-$ LINERs and Seyfert galaxies only
\label{tab4}
\end{table}

The existence of a robust anti-correlation between \gam\ 
and \edd\ offers an alternative X-ray-based method to 
estimate \mbh\ in low-accreting BHs. 
Since $L_{\rm Edd}$ is a linear function of \mbh, 
one can solve the equation for \mbh, and hence constrain it by 
plugging the values of \gam\ and \lx\ as well as the intercept and 
the slope of the anti-correlation: 
\begin{equation}
\log(M_{BH})=\log(L_{\mathrm{X}})-38.11-\left[\frac{\Gamma-B}{A}\right]
\label{equ2}
\end{equation}
where A is the best-fit slope, B is the best-fit y-intercept, and
the constant 38.11 comes from the definition $L_{\mathrm{Edd}}=1.3\times10^{38}
(M_{BH}/M_{\odot})$ \ergs.

\subsection{\mbh\ Computation}
We estimated \mbh\ for the sources in Table \ref{tab1} 
(except for 8 Compton-thick candidates) using 
the Equation 1 with the best-fitting parameters corresponding 
all AGN classes for both Case 1 (spectral results from the PL model) 
and Case 2 (spectral results from more complex models). 

The \mbh\ values for the 47 LLAGNs from 
the anti-correlation of \gam$-$\edd\ (\mbhx) and 
the ratio between the \mbhx\ and the corresponding values 
determined with dynamical methods (\mbhd) are reported 
in Table \ref{tab5}, with columns 2 and 3 referring to Case 1 
and columns 4 and 5 to Case 2.  
The uncertainty of \mbhx\ was derived from the parameter's 
uncertainty in Equation (\ref{equ2}). 

\begin{table}
\footnotesize
\begin{adjustwidth}{-1.4cm}{}
\caption{$M_{\rm BH}$ Estimation of LLAGN}
\begin{center}
\footnotesize
\begin{tabular}{lcccc}    
\hline        
\hline
\noalign{\smallskip}
LLAGN & $\log(M_{\rm BH,X})$ & $\log(\frac{M_{\rm BH,X}}{M_{\rm BH,dyn}})$ 
& $\log(M_{\rm BH,X})$ & $\log(\frac{M_{\rm BH,X}}{M_{\rm BH,dyn}})$ \\
\multicolumn{1}{c}{(1)} & (2) & (3) & (4) & (5)\\
\hline
\noalign{\smallskip}
\multicolumn{5}{c}{LINER}\\ 
\noalign{\smallskip}
\hline
I 1459 & $ 9.78 \pm 0.02 $ & $ 0.34 \pm 0.20 $ & $ 9.56 \pm 0.10 $ &$ 0.12 \pm 0.22 $\\
I 4296 & $ 8.86 \pm 0.38 $ & $ -0.27 \pm 0.38 $ & \\       
N 224 & $ 7.98 \pm 0.22 $ & $ -0.19 \pm 0.27 $ & \\       
N 2748 & $ 8.52 \pm 1.79 $ & $ 0.85 \pm 1.85 $ & \\       
N 2787 & $ 8.25 \pm 0.07 $ & $ 0.61 \pm 0.09 $ & $ 8.07 \pm 0.25 $ &$ 0.43 \pm 0.25 $\\
N 3031 & $ 9.54 \pm 0.29 $ & $ 1.64 \pm 0.30 $ & $ 7.89 \pm 0.03 $ &$ -0.01 \pm 0.10 $\\
N 3608 & $ 9.14 \pm 0.44 $ & $ 0.82 \pm 0.47 $ & \\       
N 3998 & $ 9.90 \pm 0.16 $ & $ 1.53 \pm 0.46 $ & $ 9.60 \pm 0.19 $ &$ 1.23 \pm 0.47 $\\
N 4261 & $ 5.99 \pm 1.01 $ & $ -2.75 \pm 1.01 $ & $ 8.84 \pm 0.58 $ &$ 0.10 \pm 0.58 $\\
N 4278 & $ 9.53 \pm 0.29 $ & $ 0.33 \pm 0.29 $ & $ 9.38 \pm 0.02 $ &$ 0.18 \pm 0.02 $\\
N 4459 & $ 8.39 \pm 0.30 $ & $ 0.52 \pm 0.31 $ & $ 8.17 \pm 0.43 $ &$ 0.30 \pm 0.44 $\\
N 4486 & $ 9.31 \pm 0.18 $ & $ -0.25 \pm 0.22 $ & \\       
N 4596 & $ 6.64 \pm 1.26 $ & $ -1.28 \pm 1.27 $ & $ 7.77 \pm 1.38 $ &$ -0.15 \pm 1.39 $\\
N 5077 & $ 8.60 \pm 0.87 $ & $ -0.30 \pm 0.90 $ & \\														            
\hline
\noalign{\smallskip}
\multicolumn{5}{c}{Seyfert}\\
\noalign{\smallskip}
\hline
N 1399 & $ 12.32 \pm 3.68 $ & $ 3.61 \pm 3.68 $ &        \\
N 3227 & $ 9.11 \pm 0.33 $ & $ 1.93 \pm 0.40 $ & $ 7.70 \pm 0.66 $ &$ 0.52 \pm 0.70 $\\
N 4026 & $ 7.80 \pm 1.94 $ & $ -0.53 \pm 1.95 $ & $ 8.28 \pm 2.34 $ &$ -0.05 \pm 2.34 $\\
N 4151 & $ 3.72 \pm 4.88 $ & $ -3.93 \pm 4.88 $ & $ 8.48 \pm 0.73 $ &$ 0.83 \pm 0.73 $\\
N 4258 & $ 9.85 \pm 0.88 $ & $ 2.27 \pm 0.88 $ & $ 8.02 \pm 2.00 $ &$ 0.44 \pm 2.00 $\\
N 4395 & $ 5.55 \pm 0.79 $ & $ 0.51 \pm 0.79 $ & $ 4.56 \pm 1.13 $ &$ -0.48 \pm 1.13 $\\
N 4594 & $ 9.06 \pm 0.04 $ & $ 0.30 \pm 0.41 $ & $ 8.77 \pm 0.09 $ &$ 0.01 \pm 0.42 $\\
N 5128 & $ 2.45 \pm 4.44 $ & $ -6.03 \pm 4.44 $ & $ 3.91 \pm 2.96 $ &$ -4.57 \pm 2.96 $\\
N 5252 & $ 9.07 \pm 0.66 $ & $ 0.07 \pm 0.75 $ & \\
N 6251 & $ 11.49 \pm 0.12 $ & $ 2.71 \pm 0.19 $ & $ 10.88 \pm 0.07 $ & $ 2.10 \pm 0.16 $\\
N 7457 & $ 5.48 \pm 1.67 $ & $ -1.13 \pm 1.68 $ & $ 6.01 \pm 1.82 $ &$ -0.60 \pm 1.83 $\\
\hline
\noalign{\smallskip}
\multicolumn{5}{c}{Unclassified}\\
\noalign{\smallskip}
\hline
N 221 & $ 6.38 \pm 0.19 $ & $ -0.11 \pm 0.21 $ & $ 6.41 \pm 0.15 $ &$ -0.08 \pm 0.18 $\\
N 821 & $ 7.08 \pm 1.97 $ & $ -0.55 \pm 1.97 $ & \\       
N 1023 & $ 8.20 \pm 0.04 $ & $ 0.54 \pm 0.06 $ & \\       
N 1300 & $ 3.69 \pm 1.75 $ & $ -4.16 \pm 1.77 $ & $ 9.55 \pm 0.01 $ &$ 1.70 \pm 0.29 $\\
N 2778 & $ 8.48 \pm 1.54 $ & $ 1.27 \pm 1.57 $ & \\       
N 3115 & $ 8.48 \pm 0.93 $ & $ -0.50 \pm 0.95 $ & $ 8.78 \pm 0.57 $ &$ -0.20 \pm 0.60 $\\
N 3377 & $ 8.17 \pm 1.76 $ & $ 0.11 \pm 1.77 $ & \\       
N 3384 & $ 7.25 \pm 4.40 $ & $ 0.00 \pm 4.40 $ & \\       
N 3585 & $ 9.17 \pm 0.09 $ & $ 0.64 \pm 0.15 $ & \\       
N 4291 & $ 8.56 \pm 0.47 $ & $ 0.05 \pm 0.58 $ & \\       
N 4342 & $ 7.68 \pm 2.28 $ & $ -0.88 \pm 2.29 $ & $ 7.66 \pm 0.63 $ &$ -0.90 \pm 0.66 $\\
N 4473 & $ 7.16 \pm 2.18 $ & $ -0.95 \pm 2.21 $ & \\       
N 4564 & $ 7.27 \pm 1.22 $ & $ -0.57 \pm 1.22 $ & \\       
N 4649 & $ 10.99 \pm 0.73 $ & $ 1.66 \pm 0.74 $ & $ 8.08 \pm 0.29 $ &$ -1.25 \pm 0.31 $\\
N 4697 & $ 6.09 \pm 1.12 $ & $ -2.20 \pm 1.12 $ & $ 8.35 \pm 2.72 $ &$ 0.06 \pm 2.72 $\\
N 4742 & $ 7.39 \pm 0.60 $ & $ 0.21 \pm 0.62 $ & $ 7.11 \pm 1.14 $ &$ -0.07 \pm 1.15 $\\
N 5576 & $ 7.91 \pm 3.42 $ & $ -0.35 \pm 3.42 $ & \\       
N 5845 & $ 8.83 \pm 0.72 $ & $ 0.37 \pm 0.75 $ & \\       
N 7052 & $ 11.72 \pm 0.03 $ & $ 3.12 \pm 0.22 $ & $ 8.67 \pm 1.29 $ & $ 0.07 \pm 1.31 $\\
N 4486A & $ 4.32 \pm 1.46 $ & $ -2.81 \pm 1.47 $ & $ 4.32 \pm 1.46 $ & \\
\hline
\hline
\end{tabular}
\end{center}
\footnotesize
{\bf Note :} (1) = LLAGN name; (2) \& (3) = computed \mbh\ values with Case 1 
best-fit parameters and the corresponding ratio between \mbhx\ and \mbhd\ values;   
(4) \& (5) = computed \mbh\ values with Case 2 
best-fit parameters and the corresponding ratio between \mbhx\ and \mbhd\ values
\label{tab5}
\end{adjustwidth}
\end{table}

\begin{figure}
\includegraphics[scale=0.5]{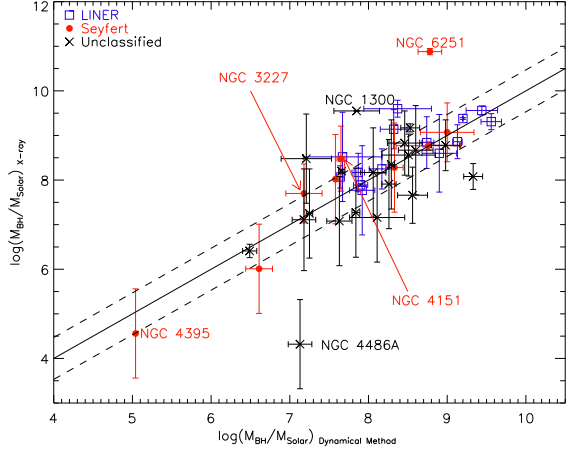}
\caption{\mbh\ values obtained using the \GLedd\ anti-correlation parameters vs. 
dynamically measured \mbh\ values. The open squares are used 
to indicate LINERs, filled circles for Seyferts, and X marks for unclassified 
sources. The one-to-one correlation is represented by the solid line whereas 
the dashed lines indicate the uncertainty.}
\label{fig7}
\end{figure}

\begin{figure}
\includegraphics[scale=0.5]{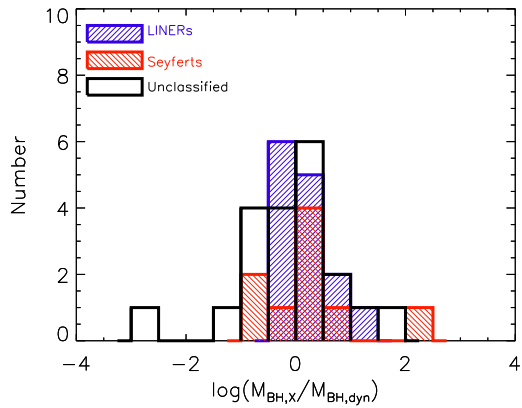}
\caption{Histogram of $\log(M_{\rm BH,X}/M_{\rm BH,dyn})$ where \mbhx\ 
refers to Case 2 (spectral results from more complex models). The 
histogram filled with positive slopes indicate LINERs, the negative 
slopes filled one is used for Seyferts, and the empty one for 
unclassified AGNs.}
\label{fig8}
\end{figure}

With few exceptions, we found a good agreement
between the \mbh\ values determined with this anti-correlation
and their corresponding dynamical values. These findings
are illustrated in Figure \ref{fig7} where we plot the \mbh\ values 
obtained with these two methods along the  y- and x-axis,
respectively. The apparent visual correlation is formally
confirmed by the statistical analysis performed using 
the {\it mpfitexy} routine 
(Markwardt 2009; Williams, Bureau \& Cappellari 2010).
The best-fit parameters, the slope and intercept, with
their $1\sigma$ uncertainty and the RMS from the one-to-one correlation
for each LLAGN class and for the combination of
all are reported in Table 6. 
The distribution of the ratio between computed \mbhx\ and its corresponding \mbhd\ 
for Case 2 is shown Figure \ref{fig8}. 

We also investigated whether X-ray radio-loudness plays a role in the 
mass determination. To this end, we have divided our sample between radio-quiet 
and radio-loud objects using as the threshold $\log R_{\rm X}\ge-2.8$ 
(Panessa et al. 2007). The values of $\log(M_{\rm BH,X}/M_{\rm BH,dyn})$ 
for radio-quiet and radio-loud objects are respectively $0.29\pm0.54$ and 
$0.55\pm1.17$, which are consistent within the errors. This suggests that 
radio-loudness does not affect the mass determination with this X-ray method.

\begin{table}
\footnotesize
\begin{adjustwidth}{0.0cm}{}
\caption{$M_{\mathrm{BH}}$ Correlation Analysis Results}
\begin{center}
\footnotesize
\begin{tabular}{lcccc} 
\hline        
\hline
\noalign{\smallskip}
\multicolumn{1}{c}{Class} & Slope & Y-int. & Spearman(Prob.) & RMS\\
\multicolumn{1}{c}{(1)} & (2) & (3) & (4) & (5)\\
\hline
\noalign{\smallskip}
\multicolumn{5}{c}{Case 1}\\
\hline
LINER & $0.56\pm 0.46$ & $ 4.06\pm 3.96$ & 0.45($1.06\times10^{-1}$) & 1.09\\
Seyfert & $1.16\pm 0.36$ & $-0.15\pm 2.82$ & 0.36($2.72\times10^{-1}$) & 2.75\\
Unclassified & $1.70\pm 0.41$ & $-5.41\pm 3.32$ & 0.64($2.37\times10^{-3}$) & 1.54\\
ALL & $1.08\pm 0.22$ & $-0.23\pm 1.80$ & 0.55($9.53\times10^{-5}$) & 1.81\\
\hline
\multicolumn{5}{c}{Case 2}\\
\hline
LINER & $0.77\pm 0.45$ & $ 2.12\pm 3.89$ & 0.72($3.48\times10^{-3}$) & 0.50\\
Seyfert & $1.38\pm 0.37$ & $-2.52\pm 2.94$ & 0.97($2.16\times10^{-5}$) & 0.83\\
Unclassified & $0.91\pm 0.38$ & $ 0.68\pm 3.05$ & 0.61($4.37\times10^{-3}$) & 0.93\\
ALL & $1.00\pm 0.23$ & $ 0.23\pm 1.89$ & 0.74($1.50\times10^{-8}$) & 0.79\\
\hline
\hline
\end{tabular}
\end{center}
\footnotesize
{\bf Note :}   
(1) AGN Class; (2) best-fit slope; (3) best-fit intercept; 
(4) Spearman's $\rho-$rank and its following probability; 
(5) RMS from the one-to-one correlation
\end{adjustwidth}
\label{tab6}
\end{table}

In summary, we computed \mbh\ for 47 LLAGNs 
based on the anti-correlation between \gam$-$\edd.
The vast majority of the \mbh\ values are in 
good agreement with their dynamical values 
within a factor of $5-6$ (RMS $\sim0.8$).

\section{Discussion}
In this work, we performed a systematic and homogeneous 
re-analysis of the X-ray spectra for a sample of LLAGNs with 
\mbh\ dynamically constrained with the aim to test the validity 
and the limitations of two X-ray-based methods to determine 
\mbh. The first method is based on the scale-invariance of 
X-ray spectral properties of BHs at all scales, whereas the 
second one is based on the anti-correlation of \gam\ vs. \edd\ 
at very low accretion rates.

\subsection{X-ray Scaling Method}
In our recent work, we demonstrated that the X-ray scaling 
method, developed for and tested on GBHs 
(Shaposhnikov \& Titarchuk 2009), can be successfully 
applied to AGNs with moderate/high accretion rate (Gliozzi et al. 2011).
Specifically, using self-similar spectral patterns 
from different GBH reference sources 
we derived the \mbh\ values of a selected sample of bright 
AGNs and then compared them with the corresponding 
values obtained from the reverberation mapping technique. 
The tight correlation found in the $\log(M_{\rm BH,scal})$ and 
$\log(M_{\rm BH,RM})$ plane (RMS $=0.35$ for the most reliable reference 
source, GRO J1655-40, and RMS$_{\rm avg}\,=0.53$ obtained by taking the 
average of five different reference patterns) 
demonstrates that the values of \mbh\ obtained with the 
scaling method are fully consistent with the reverberation 
mapping results within the respective uncertainties.

In this work, we have tested the limits of applicability of
this scaling method to low accreting AGNs with typical 
\edd\ ratio of the order of $10^{-6}-10^{-7}$, which 
correspond to very low Eddington ratios ($<10^{-4}$) for
any reasonable bolometric correction. In our starting
sample, only three objects have \edd\ that are not
extremely low: NGC 3227, NGC 4151, and NGC 4395 (all
have \edd\ $\sim10^{-3}$ and thus $L_{\rm bol}/L_{\rm Edd}\sim10^{-2}$). 
The resulting \mbh\ values derived for the LLAGN sample
from the X-ray scaling method are systematically
lower than the dynamically inferred values by three or
four orders of magnitude, indicating that the X-ray 
scaling method cannot be utilized for BHs in the very low
accreting regime. The only notable exceptions are
NGC 3227, NGC 4151, and NGC 4395 for which the derived
\mbh\ are consistent with the dynamical values.

Paradoxically, the apparent failure of the scaling
method, when applied to AGNs accreting at very low accretion 
rates, provides indirect support to this
method.  Indeed, it demonstrates that the agreement
between \mbh\ values determined with the scaling method
and the reverberation mapping values is not obtained
by chance but is based on a common spectral evolution
(the steeper when brighter spectral trend), which is systematically seen
in highly-accreting AGNs and GBHs in their transition
between the low-hard state and the soft-high state. 
This conclusion is further reinforced by the agreement
between $M_{\rm BH,scal}$ and their dynamical values that was
obtained for NGC 3227, NGC 4151, and NGC 4395, the only
sources of this sample with accretion rate close to
$10^{-2}$.

\begin{figure*}
\includegraphics[scale=0.5]{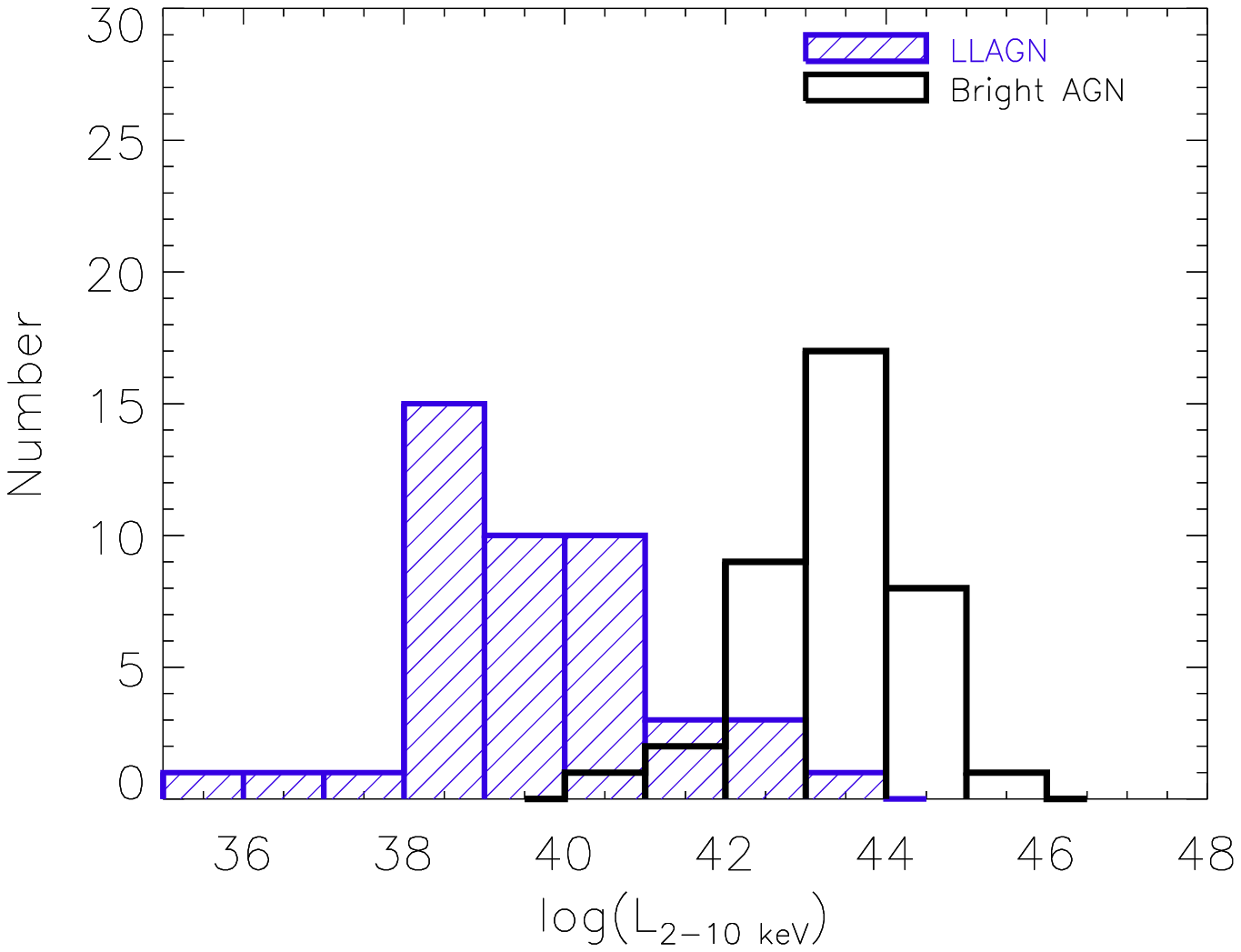}
\includegraphics[scale=0.5]{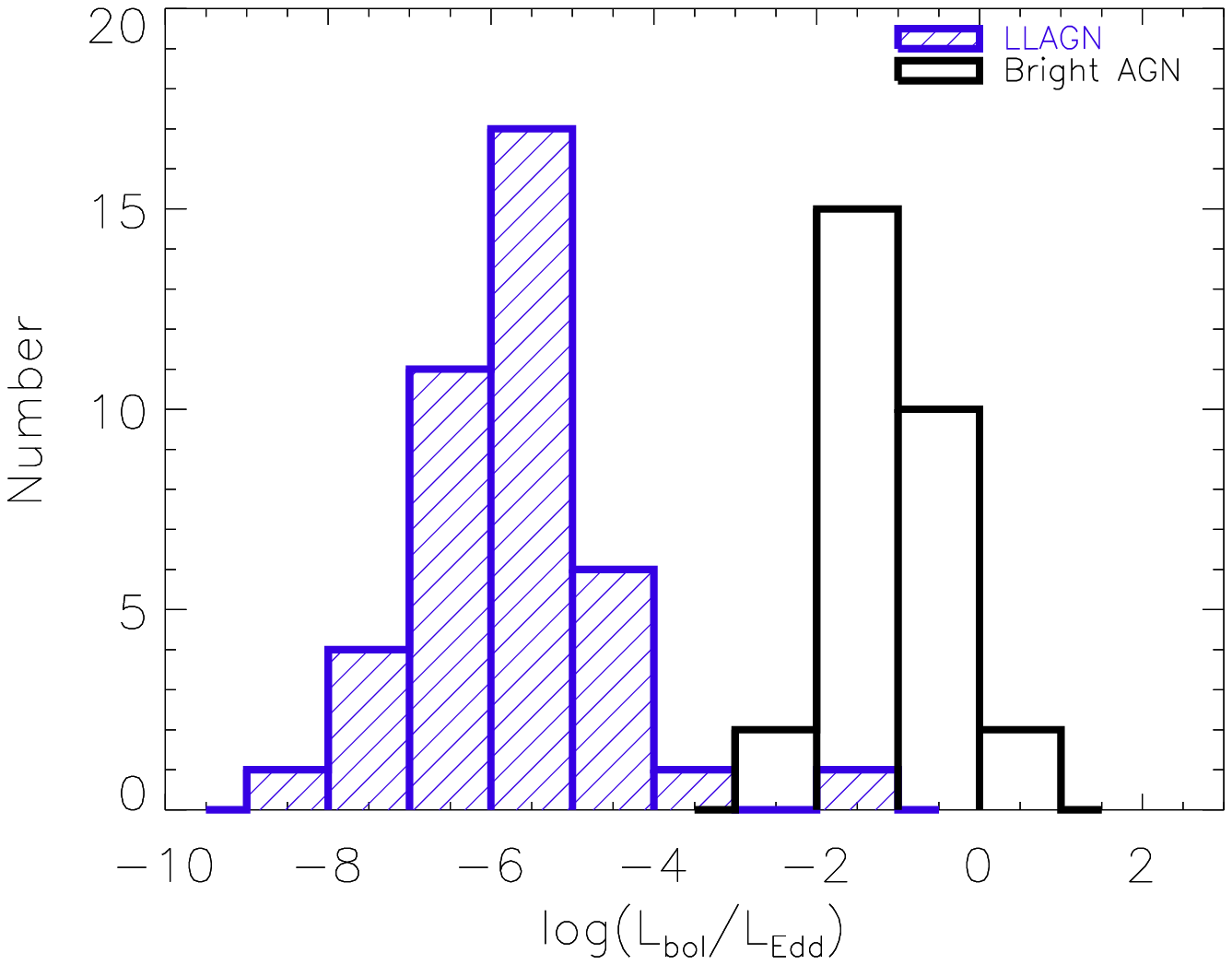}
\caption{Histogram of $\log(L_{\rm X})$ and $\log(L_{\rm bol}/L_{\rm Edd})$ for our sample of 
LLAGNs and reverberation mapping AGNs in Gliozzi et al. (2011). 
The histogram filled with negative slopes indicates 
LLAGNs and the empty one for bright AGNs in both panels.}
\label{FIG6}
\end{figure*}

In Figure \ref{FIG6} we show the histogram of
the X-ray luminosity (left panel) and of $L_{\rm bol}/L_{\rm Edd}$ 
(right panel) for the reverberation mapping sample
used by Gliozzi et al. (2011) and the LLAGN sample
utilized in the present work. The two distributions
appear to be distinct as formally demonstrated by a
Kolmogorov-Smironv (K-S) test that yields a probability
of $1.2\times10^{-13}$ and $1.7\times10^{-15}$ that the two \lx\ 
and the two $L_{\rm bol}/L_{\rm Edd}$ distributions are drawn
from the same populations. 
These combined findings
suggest that the X-ray scaling method provides reliable
estimates of \mbh\ for moderately/highly accreting
AGNs with \lx\ $>10^{42}$ \ergs\ and $L_{\rm bol}/L_{\rm Edd}\,^>_{\sim}10^{-3}$.

\subsection{Inverse Correlation of \gam\ vs. \edd}
Since LLAGNs represent the vast majority of AGNs and
many of them have X-ray observations it is important
to find an alternative way to constrain \mbh\ in
these systems exploiting their X-ray properties. Recent
studies of large samples of LLAGNs with X-ray data have 
provided solid evidence in favor of
this anti-correlation in the \gam$-$\edd\ diagram 
(e.g., \Gu\ et al. 2009; Constantin et al. 2009; Gu \& Cao 2009), 
which has been recently confirmed in an individual
LLAGN monitored for several years (Emmanoulopoulos et al. 2012).

Before comparing the results from our work to similar 
studies in the literature, it is important to underscore
the differences of these studies. In this paper, we have performed
a thorough and systematic analysis of the highest quality
spectra available for a sizable sample of LLAGNs with \mbh\
dynamically determined. Starting with a simple power-law
absorption model we progressively increased the complexity
of the spectral model to account for partial absorption,
thermal emission, soft excess, reflection, and
emission lines when necessary. In this way, the vast
majority of the sources yielded photon indices in the 
physical range from 1-3 that followed the \gam$-$\edd\ 
anti-correlation.

\Gu\ et al. (2009) used a similar sample but limited
themselves to Chandra data in the $2-10$ keV and used a relatively
simple spectral model that for some
sources yielded negative photon indices and for other values
steeper than 3. This resulted into an anti-correlation 
described by a slope of $-0.24\pm0.12$ which is not statistically significant 
but consistent with our results when using a power-law model (Case 1).

Constantin et al. (2009) used a very large sample of LLAGNs
candidates obtained by cross-matching the SDSS catalog with 
X-ray selected sources from the \chandra\ Multiwavelength Project 
(ChaMP). With such a large sample and the relatively low number of
counts (the mean source count of the sample was 76 counts) only a basic spectral
analysis can be performed providing \gam\ values ranging from -2 to
6. The anti-correlation derived from this study combining Seyfert
galaxies, LINERs and transition objects is again consistent
with our results from the power law analysis. 

The study more similar to ours in terms of statistical significance 
of the anti-correlation, quality of the spectra and reasonable values 
of \gam\ (although with a considerably smaller sample) is the one from
Younes et al. (2011), who studied a sample of 13
LINERs with \chandra\ and \xmm\ data. Their anti-correlation $-0.31\pm0.06$ 
is fully consistent with our correlation for LINERs, but slightly steeper than the 
correlation obtained combining all classes of AGNs in Case 2. 
In summary, several previous studies based on different samples and
spectral quality have provided findings fully consistent with the
anti-correlation derived in this work. 

Note that the existence
of a positive \gam$-$\edd\ correlation has been
widely accepted for more than two decades and is generally
explained in the framework of Comptonization
models by the cooling of the corona produced by an
higher flux of soft photons caused by the increased accretion
rate in the disk. On the other hand, substantial
evidence of a negative \gam$-$\edd\ correlation has
been presented only recently and its explanation is still
a matter of debate. In the framework of Comptonization
models, this anti-correlation can be explained by
a decrease of the number of scatterings associated with
very low-accreting, low-density flows and the change of
the source of Comptonized seed photons (e.g., Esin et
al. 1997; Gardner \& Done 2012; Qiao \& Liu 2013). Alternatively,
it can be explained by the dominance of the jet emission in the X-ray
range that emerges in the very low accreting regime
(e.g., Yuan \& Cui 2005).
Independently of the physical reason, the sole presence
of this inverse trend in the \gam$-$\edd\ diagram
makes it possible to constrain \mbh, because $L_{\rm Edd}$  
is a direct function of \mbh. As a consequence, by using
the best fitting parameters of the inverse correlation and
the values of \gam\ and \lx, which are obtained from the X-ray analysis,
it is possible to determine \mbh\ for any LLAGNs.

With the parameters derived from our best-fitting
anti-correlation we derived the \mbh\ values for our sample
of LLAGNs. The vast majority of the objects have \mbh\ consistent
with the corresponding dynamical values within a factor of 10
with a substantial fraction ($26/43$) within a factor of 3.

\subsection{Summary}
In conclusion, we can summarize our main results 
as follows. 
\begin{itemize}
\item The X-ray scaling method provides \mbh\ 
values in good agreement with the corresponding 
dynamically determined values not only for BH systems 
accreting at high level (as demonstrated by the reverberation 
mapping sample) but also at moderately low level (\edd\ $\sim10^{-3}$) 
as shown by NGC 3227, NGC 4151, and NGC 4395. 

\item We have also computed the X-ray radio-loudness parameter $R_{\rm X}$ for our
sample to test whether it plays a relevant role in the $\Gamma-L_{\rm X}/L_{\rm Edd}$
anti-correlation. We found that $R_{\rm X}$ does not play any significant role
in the anti-correlation (and hence in the \mbh\ determination). This is in agreement
with the findings of Sikora et al. (2007), who found that all
AGN classes follow two similar trends (named radio-loud and radio-quiet sequences) 
when the radio loudness is plotted versus the Eddington ratio. For moderately
high values of the Eddington ratio, there is an inverse trend between $R$ and $L_{\rm bol}/L_{\rm Edd}$,
which flattens at low values of $L_{\rm bol}/L_{\rm Edd}$. Our sample, which is characterized by
very small values of $L_{\rm bol}/L_{\rm Edd}$, appears to be fully consistent with the flat part
of the trend shown by the radio-quiet sequence.

\item For very low accreting AGNs (typically, with \lx\ $<10^{42}$ \ergs\ or \edd\ $<10^{-4}$), 
the scaling method fails to properly 
constrain \mbh\ because its basic assumption (the steeper when brighter 
trend) no longer holds. 
Nevertheless, for very-low-accreting AGNs, 
to get a reasonable estimate of \mbh\ (within a factor of $\sim10$)
we can use the equation 
$\log(M_{\rm BH}) = \log(L_{\rm X}) - (\Gamma\ -B)/A-38.11$ 
(where B is the intercept and A the slope of the anti-correlation; 
their vales are provided in Table \ref{tab4}).

\item The possibility to constrain the \mbh\ in low-accreting
systems with this simple X-ray method may
have important implications for large statistical studies
of AGNs. This is because the $L/L_{\rm Edd}$ appears to play a
crucial role in defining the properties and the evolution
of AGNs. However, in current studies  $L_{\rm Edd}$ uniquely relies
on \mbh\ estimates from optically-based indirect methods, which may not
be appropriate for all AGN classes and accretion
regimes. The X-ray approach, which is based on assumptions
completely different from those used in the
optically-based indirect methods, can be thus used as
a sanity check. In
addition, it may expand the range of the investigation
of the cosmic evolution of galaxies, since it can be applied
to very low accreting systems, which represent the
majority of the AGN population.
\end{itemize}
\section*{Acknowledgments}
We thank the referee that led to a significant improvement of this paper. 
This research has made use of NED which is operated by the 
Jet Propulsion Laboratory, California Institute Technology, under 
contract with NASA.

\label{lastpage}


\end{document}